\documentclass[twocolumn]{aastex631}

\usepackage[utf8]{inputenc}
\usepackage{textcomp}

\newcommand{\HI}{H\,{\sc i}}
\shorttitle{\HI\ Halo of NGC~2683}
\shortauthors{Jiao et al.}
\graphicspath{{./}{figures/}}

\begin{document}

\title{FAST Reveals the Extended \HI\ Halo and Accretion Signatures of  NGC~2683}

\correspondingauthor{Qian Jiao, Ming Zhu}
\email{jiaoqian@whpu.edu.cn, mz@nao.cas.cn}

\author{Qian Jiao}
\affiliation{School of Electrical and Electronic Engineering, Wuhan Polytechnic University, Wuhan 430023, People's Republic of China}

\author{Ming Zhu}
\affiliation{National Astronomical Observatories, Chinese Academy of Sciences, Beijing 100101, People's Republic of China}
\affiliation{University of Chinese Academy of Sciences, Beijing 100049, People's Republic of China}
\affiliation{Guizhou Radio Astronomical Observatory, Guizhou University, Guiyang 550000, People's Republic of China}
\affiliation{CAS Key Laboratory of FAST, National Astronomical Observatories, Chinese Academy of Sciences, Beijing 100101, People's Republic of China}

\author{Bernd Vollmer}
\affiliation{Universit\'e de Strasbourg, CNRS, Observatoire Astronomique de Strasbourg, UMR 7550, 67000 Strasbourg, France}

\author{Mei Ai}
\affiliation{National Astronomical Observatories, Chinese Academy of Sciences, Beijing 100101, People's Republic of China}
\affiliation{Guizhou Radio Astronomical Observatory, Guizhou University, Guiyang 550000, People's Republic of China}
\affiliation{CAS Key Laboratory of FAST, National Astronomical Observatories, Chinese Academy of Sciences, Beijing 100101, People's Republic of China}

\author{Haiyang Yu}
\affiliation{National Astronomical Observatories, Chinese Academy of Sciences, Beijing 100101, People's Republic of China}
\affiliation{University of Chinese Academy of Sciences, Beijing 100049, People's Republic of China}

\author{Qinghua Tan}
\affiliation{Purple Mountain Observatory, Chinese Academy of Sciences, 10 Yuanhua Road, Nanjing 210023, People’s Republic of China}

\author{Cheng Cheng}
\affiliation{Chinese Academy of Sciences South America Center for Astronomy, National Astronomical Observatories, Chinese Academy of Sciences, Beijing 100101, People's Republic of China}
\affiliation{Key Laboratory of Optical Astronomy, National Astronomical Observatories, Chinese Academy of Sciences, 20A Datun Road, Chaoyang District, Beijing 100101, People's Republic of China}

\author{Yang Gao}
\affiliation{College of Physics and Electronic Information, Dezhou University, Dezhou 253023, People’s Republic of China}



\begin{abstract}

We present the results of our recent \HI\ observations conducted on the edge-on galaxy NGC~2683 using the Five-hundred-meter Aperture Spherical radio Telescope (FAST). In comparison to previous observations made by the VLA, FAST has detected a more extensive distribution of \HI. Particularly noteworthy is that the detections made by FAST extend approximately four times farther than those of the VLA in the vertical direction from the galactic plane. The total \HI\ flux measured for NGC~2683 amounts to $F_{\rm HI} = 112.1\,\rm{Jy\,km\,s^{-1}}$ (equivalent to a total \HI\ mass of $M_{\rm HI} = 2.32 \times 10^9\,{\rm M_\odot}$), which is slightly higher than that detected by VLA. FAST has also identified three dwarf galaxies in close proximity to NGC~2683, namely KK~69, NGC2683dw1 (hereafter dw1), and NGC2683dw3$?$ (hereafter dw3$?$). dw3$?$ is situated within the extended \HI\ distribution of NGC~2683 in projection and lies near the tail of KK~69 extending towards NGC~2683. These observations suggest that dw3$?$ is likely a result of the accretion process from NGC~2683 to KK~69. Furthermore, FAST has detected three high-velocity clouds (HVCs), with complex B potentially undergoing accretion with NGC~2683. Based on the model from \citet{vollmer2015flaring} and incorporating the \HI\ halo component, we found that the model with the added \HI\ halo aligns more closely with our FAST observations in NGC~2683. The estimated mass of this \HI\ halo is $3 \times 10^8\,{\rm M_\odot}$, constituting approximately 13\% of the total \HI\ mass of the galaxy. We suggest that the origination of this \HI\ halo is more likely attributed to external gas accretion.

\end{abstract}



\section{Introduction} \label{sec:intro}


NGC~2683, classified as an edge-on Sb galaxy with $M_{\rm B} = -19.59$\,mag, has been the subject of observational studies across multiple bands (e.g., \citealt{barbon1975photographic, harris1985globular, proctor2008keck, de2009kinematic, vollmer2015flaring, crosby2023new}). Over time, NGC~2683 has been scrutinized in \HI\ with increasing sensitivity in various research endeavors, including investigations by \citet{casertano1991declining, vollmer2015flaring, zheng2022chang}.\citet{casertano1991declining} conducted a one-hour observation of NGC~2683 using the VLA in D array configuration, revealing neutral hydrogen extending substantially beyond the visible light boundaries of the galaxy on both sides. Subsequent \HI\ observations by \citet{vollmer2015flaring} with VLA C and D arrays, characterized by enhanced sensitivity, unveiled an atomic hydrogen distribution spanning a diameter of 26.5 kpc, nearly three times the optical diameter. They attributed the vertical extent of the \HI\ distribution to the projection of a flaring gas disk.

NGC~2683 exhibits a high inclination angle, with \citet{funes2002position} reporting an inclination of 78$^{\circ}$, while other studies suggest a minimum inclination of 80$^{\circ}$ (e.g., \citealt{barbon1975photographic, broeils1994search}). Leveraging higher sensitivity \HI\ data, \citet{vollmer2015flaring} identified a thin disk inclined at 80$^{\circ}$, with the outer low surface brightness ring best modeled at an inclination of 87$^{\circ}$. The elevated inclination angle of NGC~2683 renders it a favorable candidate for investigating extraplanar diffuse gas.

NGC~2683 is situated within the Leo Spur at a distance of 9.36$\,$Mpc \citep{saponara2020new, crosby2023new}. The galaxy has a heliocentric velocity of 411$\,{\rm km\,s^{-1}}$ \citep{haynes1998asymmetry} and a virial radius of 220$\,$kpc \citep{crosby2023new}. Previous investigations have been carried out to explore satellite galaxies in the vicinity of NGC~2683, resulting in the identification and imaging of twelve candidates based on optical observations (e.g., KK~69, KK~70, N2683dw1 and N2683dw2, \citealt{karachentseva1998list, karachentsev2015peculiar, javanmardi2016dgsat, carlsten2022exploration, crosby2023new}). An additional entity, an \HI\ cloud identified as NGC2683dw3$?$ in proximity to NGC~2683, lacking an optical counterpart, was detected through radio frequency observations \citep{saponara2020new}.

Among these companion galaxies, KK~69 stands out as the brightest dwarf galaxy with an apparent magnitude of $M_{\rm B} = -12.5\,$mag. The distance of KK~69 is estimated to be 9.28$\,$Mpc, determined through the tip magnitude of the red giant branch \citep{karachentsev2015peculiar}, with a projected separation from NGC~2683 measuring 23$'$ ($\sim 62\,$kpc). Observations of the \HI\ gas in KK~69 have been conducted by various researchers such as \citet{huchtmeier2003hi, begum2008figgs, saponara2020new}. \citet{saponara2020new} utilized the Giant Metrewave Radio Telescope (GMRT) to study KK~69 and found that the \HI\ gas in KK~69 is offset with respect to its stellar body, possibly due to interactions between the \HI\ gas disk and the intragroup medium. Additionally, \citet{vollmer2015flaring} discussed the potential link between their observed flare in NGC~2683 and external accretion. These previous observations indicate the possibility that NGC~2683 may be interacting with its companion dwarf galaxies, with one of them possibly serving as the source for external accretion of \HI\ gas.

NGC~2683 is a promising candidate for investigating extraplanar gas and its interactions with nearby dwarf galaxies. In this study, we present our latest observations of the \HI\ gas within NGC~2683 and its neighboring satellites, utilizing the Five-hundred-meter Aperture Spherical radio Telescope (FAST, \citealt{nan2011five, jiang2019commissioning, jiang2020fundamental}). With its impressive 500m diameter and low system temperature (below 24$\,$K, \citealt{jiang2020fundamental}), FAST is currently recognized as the most sensitive single dish radio telescope globally.  Our research has achieved observations of significantly higher sensitivity compared to previous studies, unveiling a more extensive distribution of \HI\ gas. Furthermore, We not only detected the \HI\ distribution of three dwarf galaxies in the vicinity of NGC~2683, but also discovered three \HI\ complexes. Section 2 provides details on the observation and data reduction, and Section 3 outlines the results. In Section 4, we discuss the \HI\ halo and its potential interaction mechanisms, with Section 5 summarizing our key conclusions.

\begin{figure*}
	\includegraphics[width=0.95\textwidth]{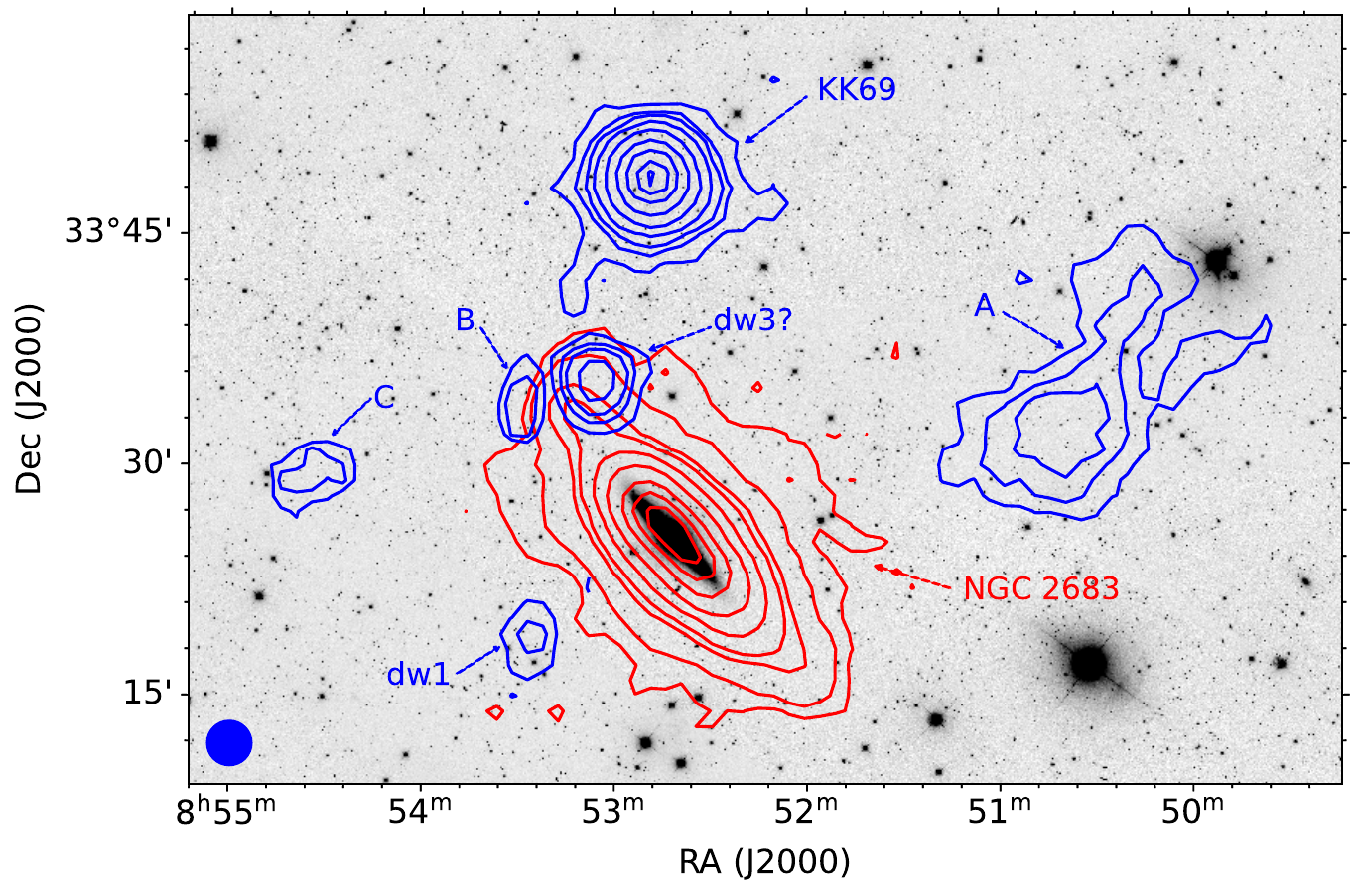}
    \caption{The \HI\ flux-density contours in red and blue colors, overlaid on the DECaLs optical image of the NGC~2683 galaxy group. The red contours represent the FAST data integrated over the velocity range of $170-650\,\rm{km\,s^{-1}}$ for the NGC~2683 galaxy, and the contour levels are (1, 3, 10, 15, 30, 60, 120, 180, 230) $\times 5\sigma$ ($5\sigma$ corresponds to $\rm{0.1\,Jy/beam}\,km\,s^{-1}$). The blue contours represent integrated FAST data of satellite galaxies and three complexes A, B, and C. For KK~69, dw1, dw3$?$, and complex A, the contour levels are (1, 2, 3, 5, 10, 20, 35, 50, 60) $\times 5\sigma$. As for complexes B and C, the contour levels are 1, and 1.5 $\times 5\sigma$. Additionally, a blue filled circle is used to indicate the beam size of $2.9'$.}
    \label{fig:moment0-optical}
\end{figure*}

\begin{table*}
	\centering
	\caption{The basic information and \HI\ properties of NGC~2683 group.}
	\label{table:regions}
	\begin{tabular}{lccccccccr} 
		\hline
		Name & RA. & Dec. & Velocity Range & $5\sigma$ in flux density & $5\sigma$ in column density & Distance & \HI\ flux & $M_{\rm HI}$ \\
		    &  & & $\rm{km\,s^{-1}}$ &  $\rm{Jy/beam\,km\,s^{-1}}$ & $10^{18}\,\rm{cm^{-2}}$ & Mpc & $\rm{Jy\,km\,s^{-1}}$ & $\rm{\times 10^7\,M_\odot}$ \\
		\hline
		NGC~2683 & 08:52:40.9 & +33:25:02 & 170-650 & 0.1 & 3.7 & 9.36 & 112.1 & 231.7 \\
		KK~69  & 08:52:50.8 & +33:47:52.0 & 420-500 & 0.04 & 1.5 & 9.28 & 4.7 & 9.6 \\
		dw3$?$ & 08:53:06 & +33:35:20 & 420-500 & 0.04 & 1.5 & 9.28 & 0.4 & 0.8 \\
		dw1 & 08:53:26.8 & +33:18:19 & 370-440 & 0.04 & 1.5 & 9.36 & 0.1 & 0.2 \\
		A & 08:50:43 & +33:34:01 & 120-190 & 0.04 & 1.4 & & 1.4 \\
		B & 08:53:32 & +33:34:01 & 100-160 & 0.03 & 1.1 & & 0.07 \\
		C & 08:54:34 & +33:28:56 & 100-160 & 0.03 & 1.1 & & 0.09 \\
		\hline
	\end{tabular}
\end{table*}

\section{Observation and data reduction} 

Using the FAST, We mapped a sky region with a R.A. ranging from $132.03< \alpha <133.98$, and a decl. within $33.07 <\delta < 34.03$ to show the distribution of \HI\ gas surrounding NGC~2683. Our observations were carried out using FAST's focal-plane 19-beam receiver system, arranged in a hexagonal array and operating in dual polarization mode, with a frequency range from 1050\,MHz to 1450\,MHz. To capture the data, we utilized the Spec(W) spectrometer, which consists of 65,536 channels covering a bandwidth of 500\,MHz for each polarization and beam with a velocity resolution of 1.67$\,{\rm km\,s^{-1}}$. The FAST \HI\ survey was conducted using the drift scan mode, and the 19-beam receiver was rotated by $23.4^\circ$ to ensure that the beam tracks were equally spaced in decl. with a spacing of $1.14'$. The half-power beamwidth (HPBW) was approximately $2.9'$ at 1.4\,GHz for each beam, and the pointing accuracy of FAST was roughly $12''$.

The flux calibration was conducted by injecting a 10\,K calibration signal (CAL) every 32 seconds for a duration of 1 second, with the aim of calibrating the antenna temperature. The data were reduced using the HIFAST data reduction pipeline, developed by \citet{jing2024hifast}, specifically designed for processing \HI\ data obtained from the FAST. Baseline correction was performed using the asymmetrically reweighted penalized least-squares algorithm (arsPLs, \citealt{baek2015baseline}). Following full calibration of the spectra, they were gridded into an image with 1-minute spacing and created as a data cube in the standard FITS format. The antenna temperature was convert to flux density using a telescope gain of approximately 16$\,{\rm K\,Jy^{-1}}$. Further detailed procedures can be found in \citet{xu2021discovery, yu2023high}. In order to construct a highly sensitive \HI\ image, the velocity resolution of the FAST data was smoothed to 3.4$\,{\rm km\,s^{-1}}$.

The spatial distribution of the satellite galaxy NGC2683dw$?$ (hereafter dw3$?$) and complex B overlaps with NGC~2683, but their differentiation can be achieved based on their respective velocities. To investigate NGC~2683 and its satellites, we have divided the observations into different regions, as outlined in Table\,\ref{table:regions}. The root mean square (rms) brightness temperature sensitivity for all regions, with the exception of region B and C (discussed in the following section), is approximately 0.5$\,{\rm mJy\,beam^{-1}}$. This sensitivity level corresponds to a column density sensitivity of 6.1  $\times 10^{16}\,{\rm atoms\,cm^{-2}}$ per channel, with a resolution of 3.4$\,{\rm km\,s^{-1}}$ and a main beam efficiency of approximately $0.8$. For complexes B and C, the sensitivity is reduced to 0.4$\,{\rm mJy\,beam^{-1}}$.

\section{Results}

\begin{figure*}
	\includegraphics[width=0.47\textwidth]{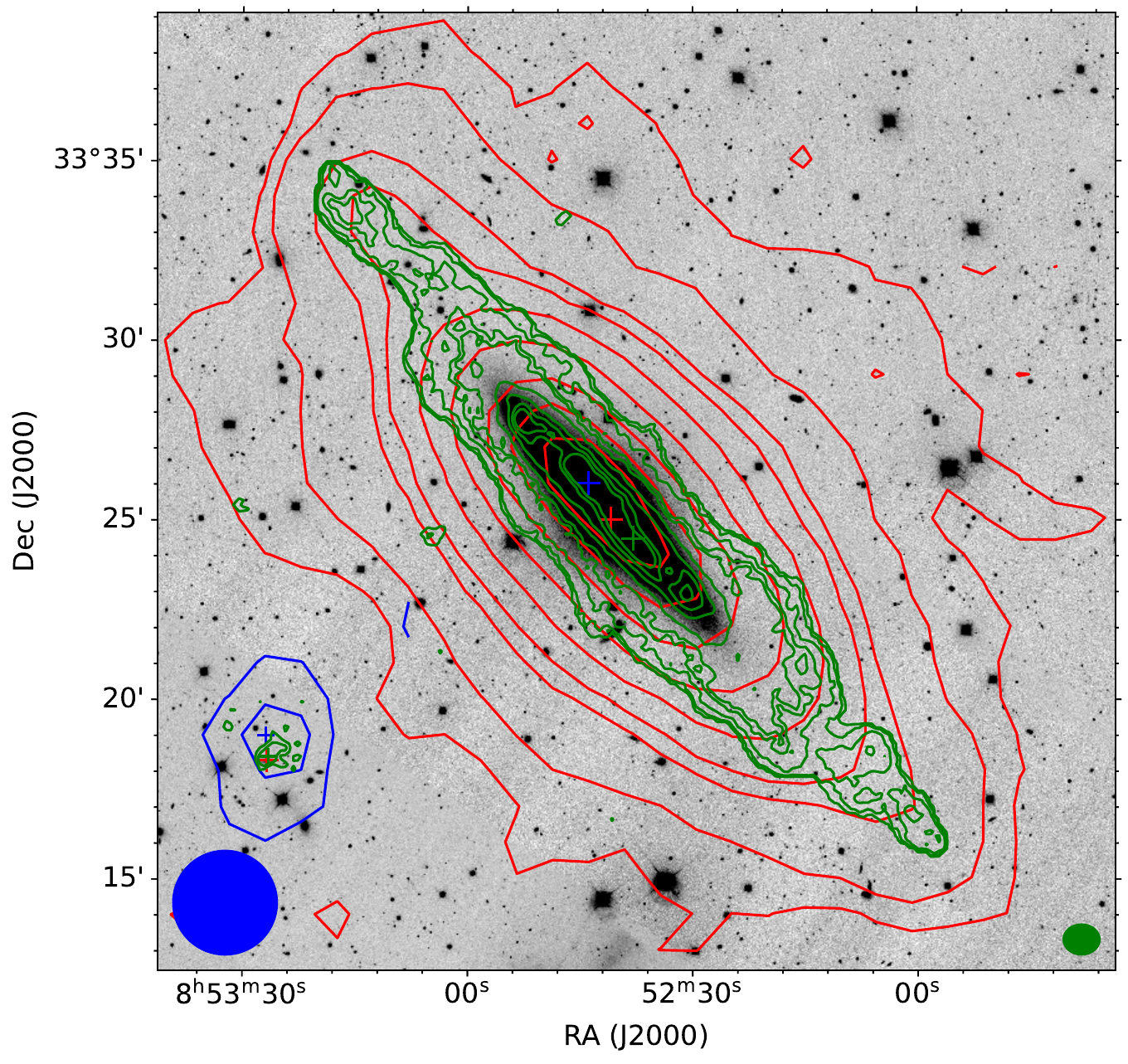}
	\includegraphics[width=0.47\textwidth]{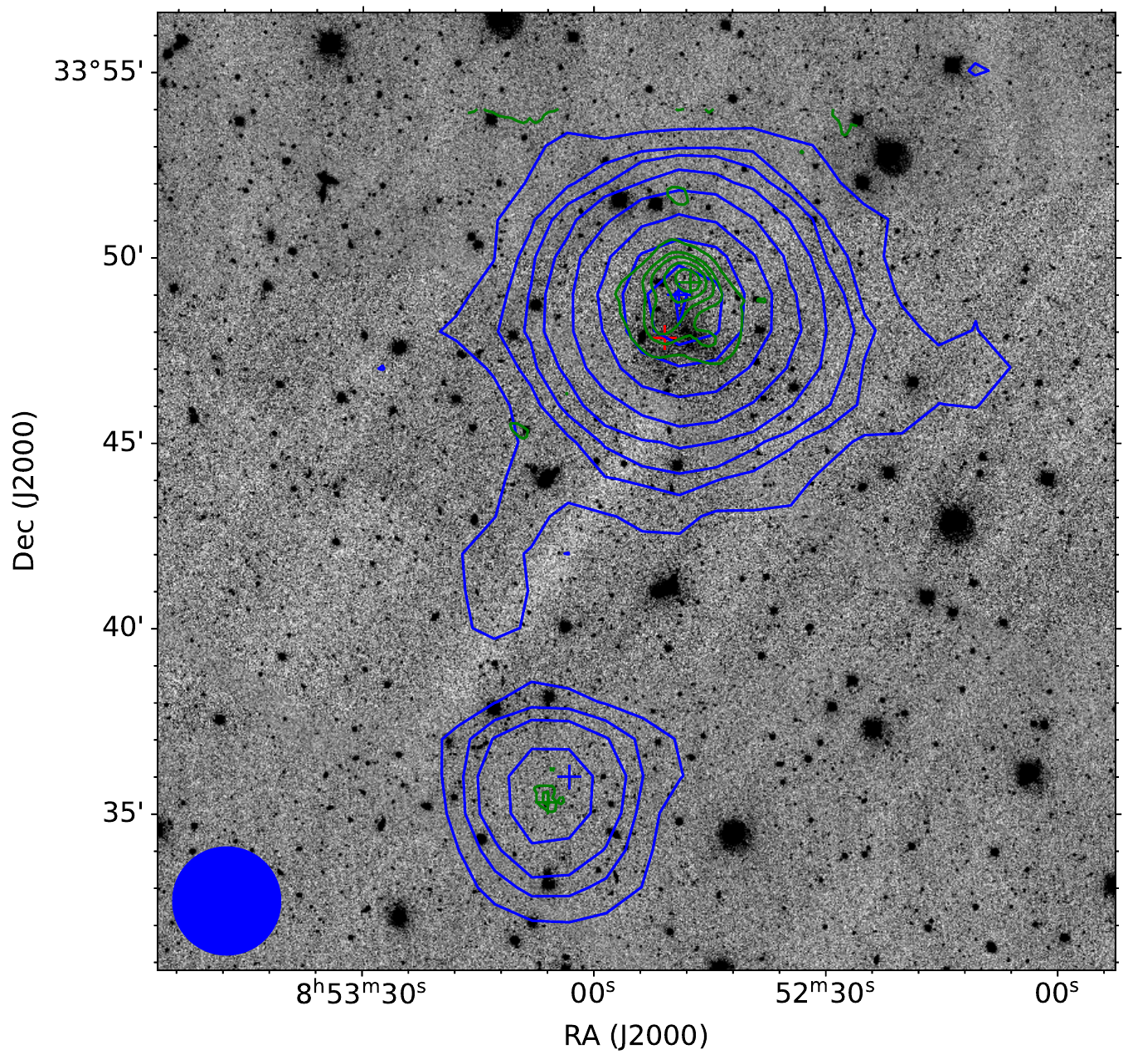}
	
    \caption{\HI\ flux-density contours overlaid on the DECaLs optical image for the NGC~2683 galaxy and dw1 in the left panel, and for KK69 and dw3$?$ in the right panel. The red and blue contours correspond to the same contours as shown in Figure\,\ref{fig:moment0-optical}, while the green contours represent VLA and GMRT observations from \citet{vollmer2015flaring, saponara2020new}. For the NGC~2683 galaxy, the contour levels for the VLA data (left panel) are set at (2, 8, 24, 48, 64, 128, 264, 392, 500) $\times \rm{10\,mJy/beam\,km\,s^{-1}}$ or $2.6\times10^{19}\,\rm{cm^{-2}}$. As for dw1 (left panel) and dw3$?$ (right panel), the contour levels for VLA observations are $4.4\times10^{19}\,\rm{cm^{-2}}$ ($5\sigma$) and $7.8\times10^{19}\,\rm{cm^{-2}}$. The green contours representing KK~69 are obtained through GMRT observations (right panel), with contour levels defined at  (1, 4, 6, 9, 12) $\times \rm{13\,mJy/beam\,km\,s^{-1}}$ or $1.5\times10^{19}\,\rm{cm^{-2}}$. The red plus symbols mark the central of stellar distribution, while the green markers indicate the peak positions of integrated \HI\ flux observed by VLA or GMRT. The blue markers represent the peak positions of integrated \HI\ flux observed by FAST. The blue filled circle indicates the beam size of $2.9'$ for FAST, while the green filled circle represents the beam size of $61'' \times 51''$ for VLA.}
    \label{fig:moment0-optical-NGC2683}
\end{figure*}

\begin{figure*}
\centering
	\includegraphics[width=0.33\textwidth]{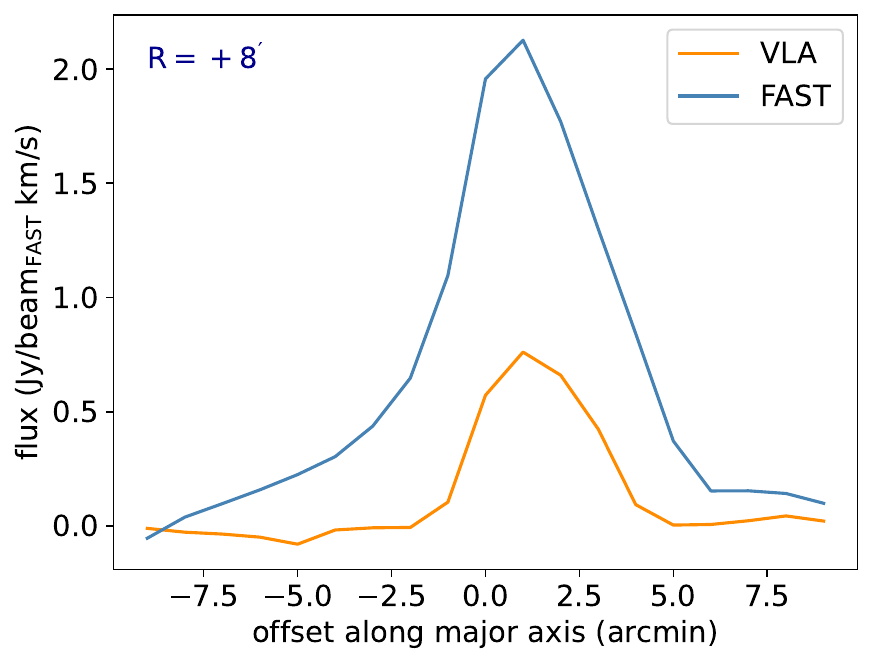}
	\includegraphics[width=0.33\textwidth]{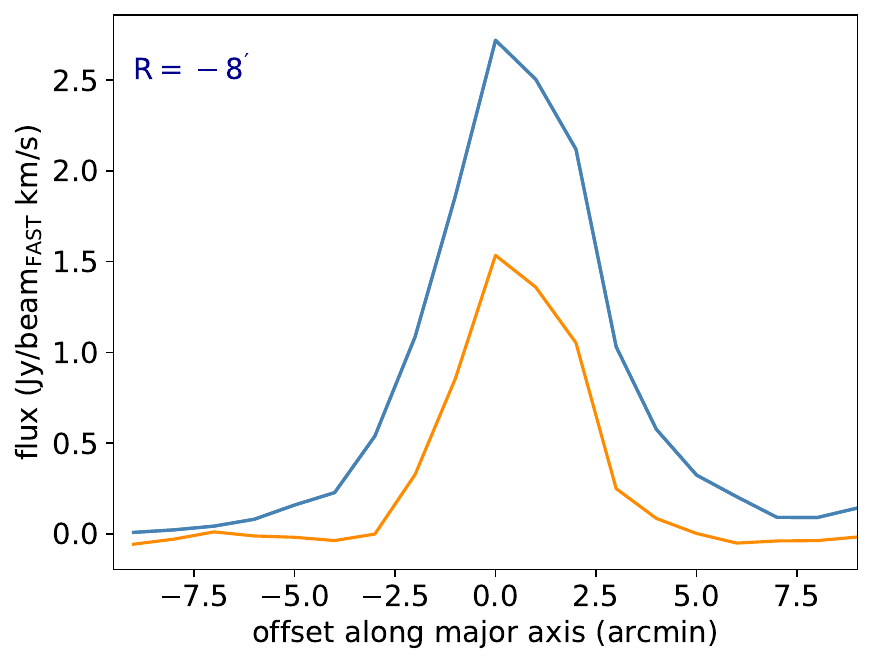}\\
		\includegraphics[width=0.33\textwidth]{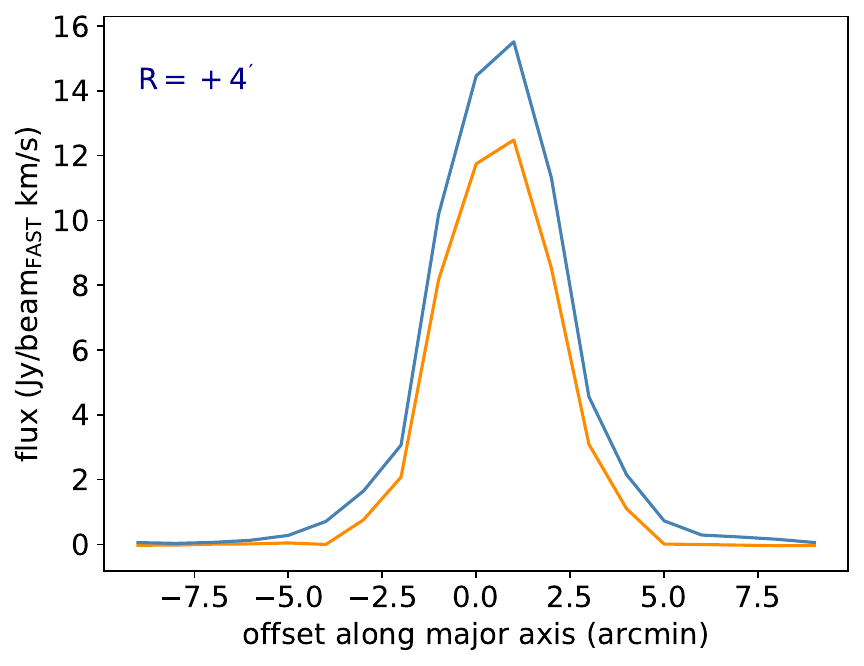}
	\includegraphics[width=0.33\textwidth]{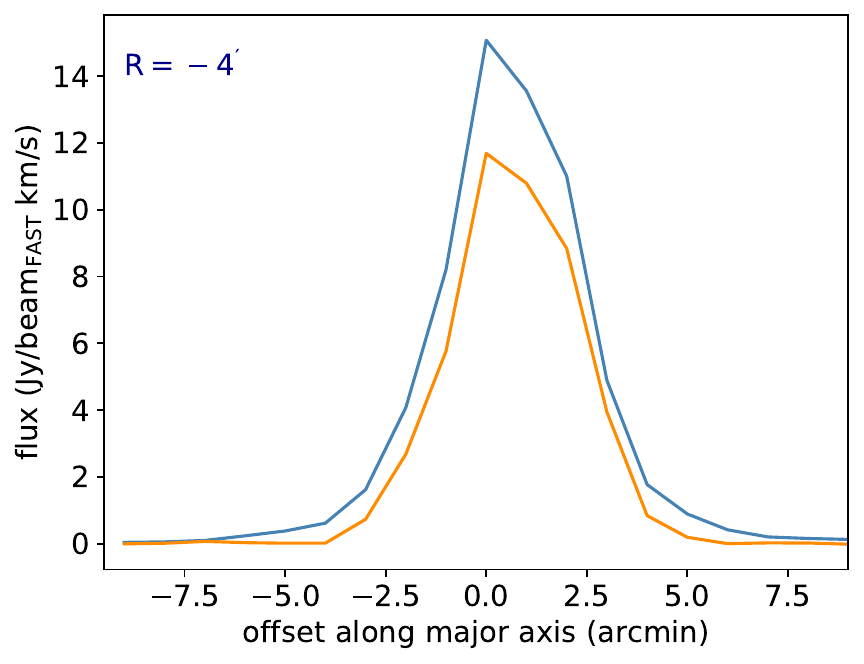}\\
	\includegraphics[width=0.33\textwidth]{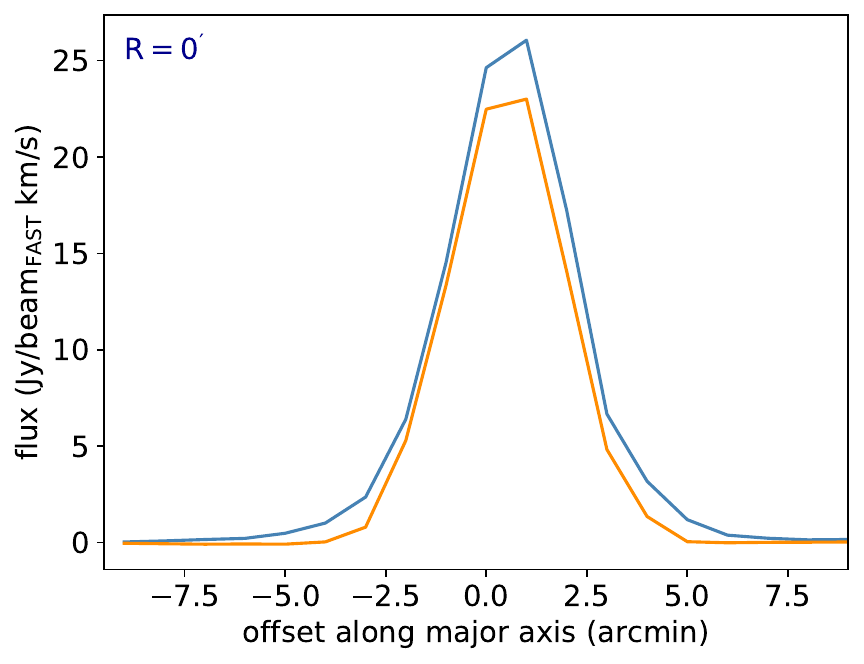}
	\includegraphics[width=0.34\textwidth]{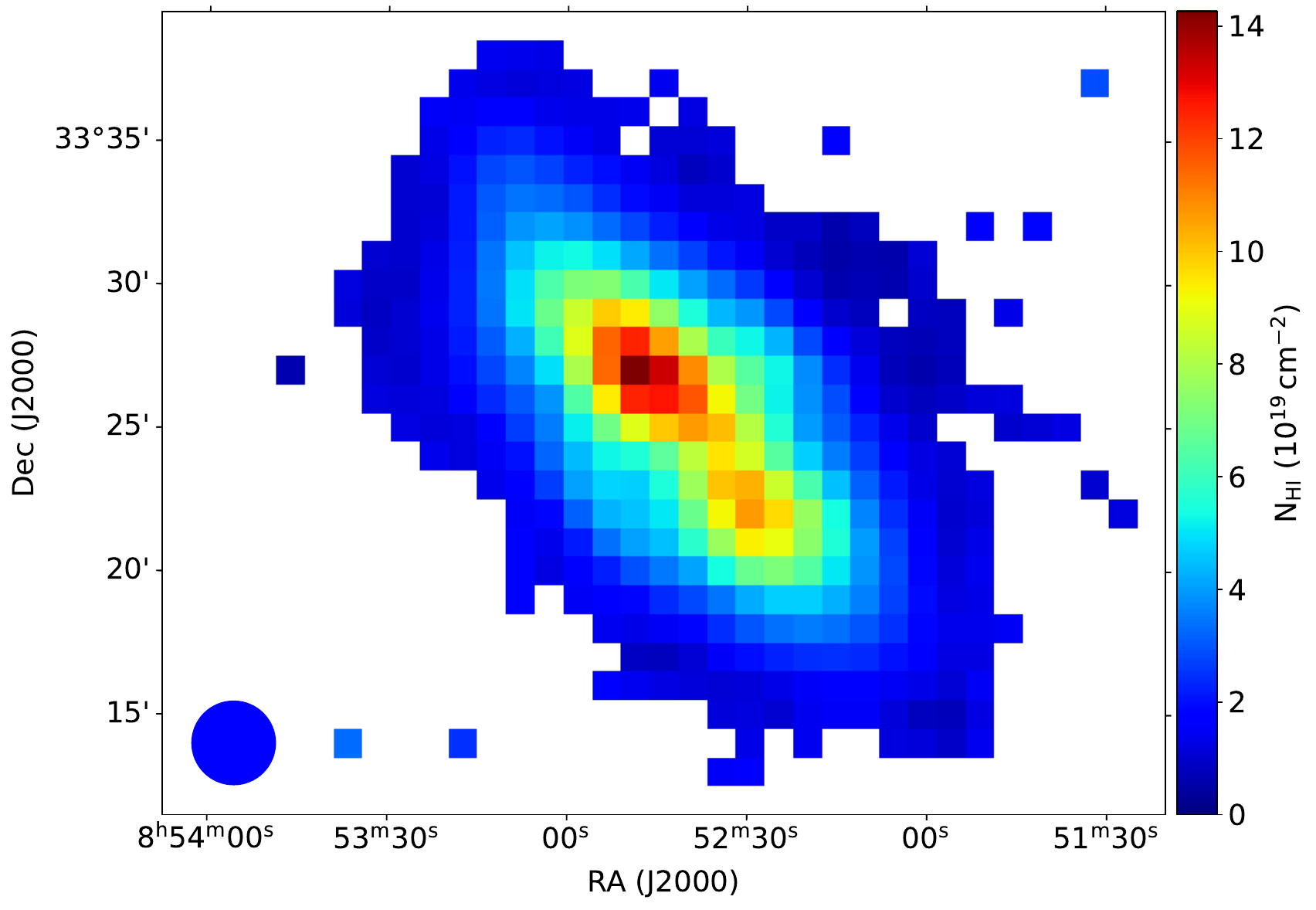}

    \caption{The left panels and the top two panels on the right respectively illustrate the flux profiles perpendicular to the galactic plane at different distances from the galactic center along the major axis. R in each panel denotes the distance from the center, with positive offsets directed towards the north-east. In this context, the steel-blue lines represent the FAST observations, while the dark-orange lines correspond to the reprocessed VLA observations. 
    The bottom right panel showcases the column density maps of the excess \HI\ detected by FAST in comparison to the VLA observations (FAST$-$VLA). The solid blue circle indicate the beam size of FAST.}
    \label{fig:compare-NGC2683}
\end{figure*}

\begin{figure*}
\centering
	\includegraphics[width=0.34\textwidth]{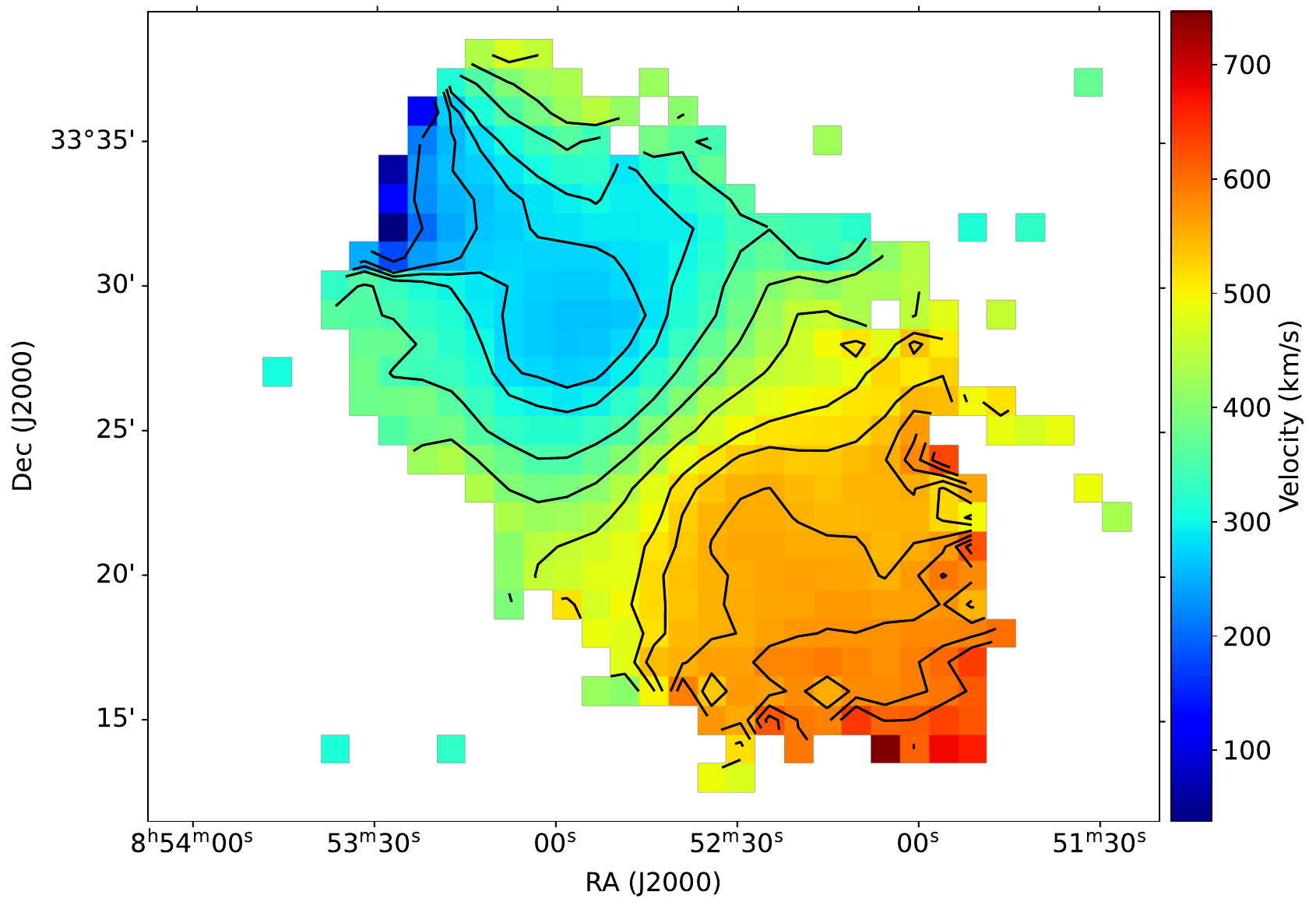}
	\includegraphics[width=0.29\textwidth]{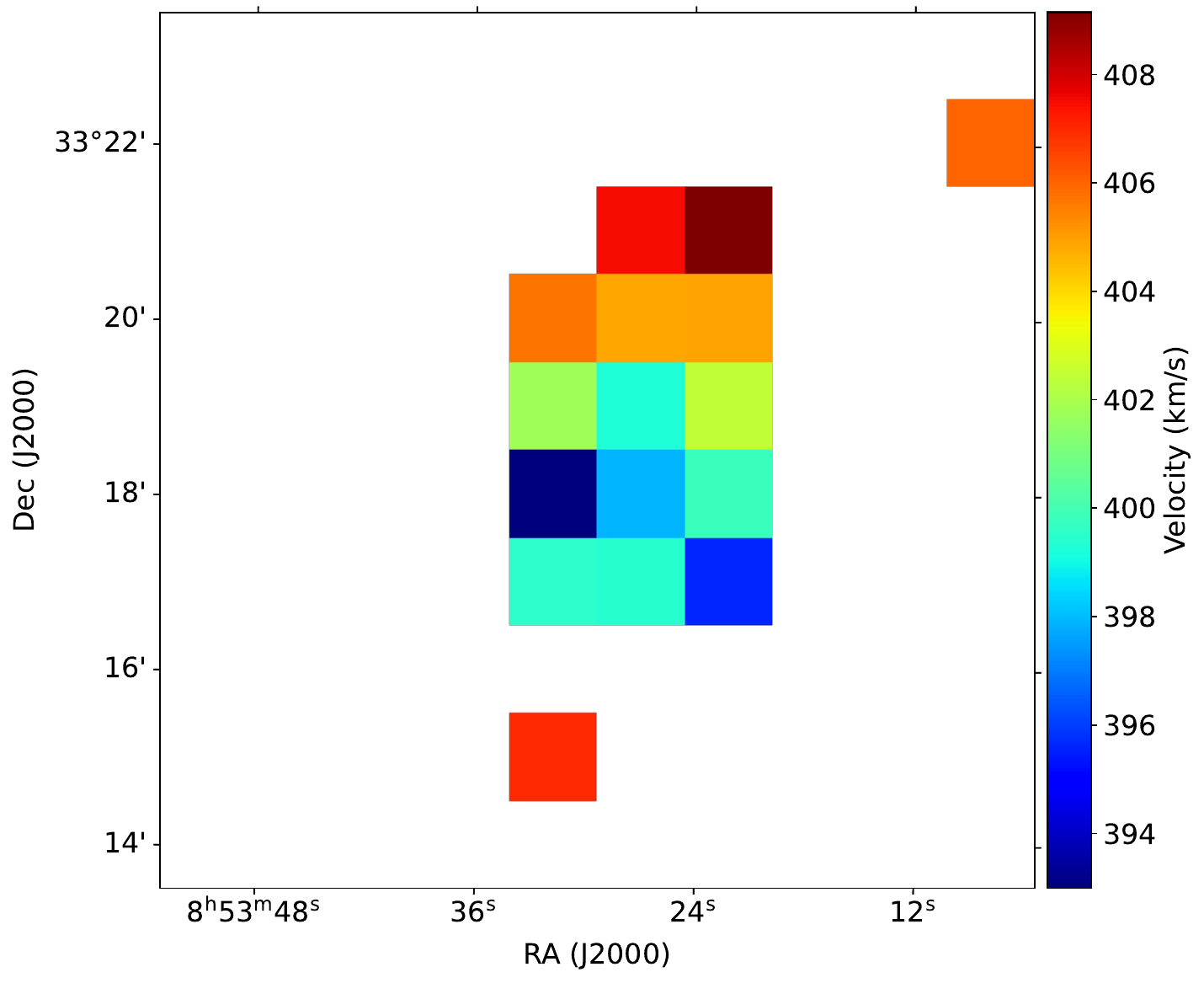}
	\includegraphics[width=0.28\textwidth]{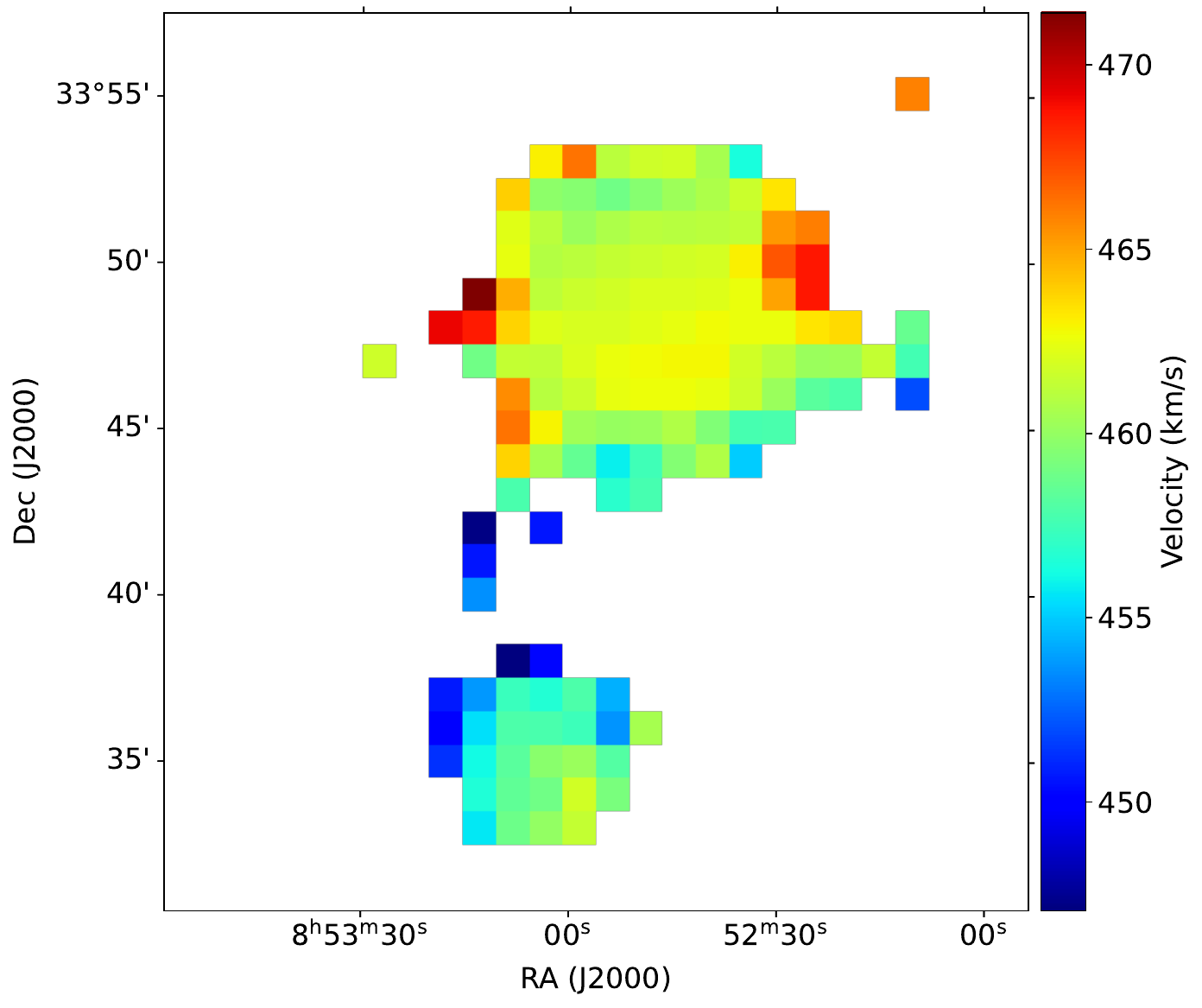}
    \caption{The velocity field derived from the \HI\ data cube, specifically the first moment map, is depicted using a color scale for the galaxy NGC~2683 (left panel), as well as its satellites dw1 (middle panel), KK~69 and dw3$?$ (right panel). The velocity contours are displayed at intervals of 200, 260, 280, 300, 350, 400, 450, 500, 525, 550, 570, 590, and 610 $\rm{km\,s^{-1}}$ for NGC~2683.}
    \label{fig:moment1-NGC2683}
\end{figure*}

Figure\,\ref{fig:moment0-optical} displays the integrated flux density contours of the \HI\ emissions of the NGC~2683 galaxy group, detected by the FAST superimposed on the optical image from the DECam Legacy Survey (DECaLs, \citealt{dey2019overview}). Notably, in addition to detecting the NGC~2683 galaxy and its satellites KK~69, NGC~2683dw1 (hereafter dw1), and dw3$?$, FAST also identified three distinct complexes denoted as A, B, and C. The integrated velocity ranges for each object are provided in Table\,\ref{table:regions}. The contours originate from a threshold of $5\sigma$ for each individual galaxy or complex, with the corresponding $5\sigma$ values listed in Table\,\ref{table:regions}. 

NGC~2683 was previously observed by \citet{vollmer2015flaring} using the VLA, which achieved a sensitivity of $3\sigma = 1.1  \times 10^{19}\,{\rm cm^{-2}}$ with a beam size of $21'' \times 20''$. In contrast, our FAST observation achieved a sensitivity of $5\sigma = 3.7  \times 10^{18}\,{\rm cm^{-2}}$, representing a significantly higher sensitivity compared to the VLA observations. The left panel of Figure\,\ref{fig:moment0-optical-NGC2683} illustrates the integrated flux density contours of NGC~2683 and dw1 from both FAST and VLA overlaid on the DECaLs optical image. Our FAST observations reveal a more extended \HI\ disk in the NGC~2683 galaxy compared to both the bright optical counterpart and VLA observations. The total \HI\ flux of the galaxy is measured to be $F_{\rm HI} = 112.1\,\rm{Jy\,km\,s^{-1}}$, which is slightly higher than the value of $F_{\rm HI} = 101.4\,\rm{Jy\,km\,s^{-1}}$ reported in \citet{vollmer2015flaring}. With a distance of 9.36$\,$Mpc, this corresponds to a total \HI\ mass of $M_{\rm HI} = 2.32 \times 10^9\,{\rm M_\odot}$.

To facilitate the comparative analysis of the data obtained from VLA and FAST, we reprocessed the raw VLA data for NGC~2683. We utilized the VLA D array data with project ID AI134. The raw data underwent calibration and cleaning using the standard calibration procedures in CASA. To match the FAST data, we applied \textit{uvtaper} in the \textit{tclean} pipeline, ultimately obtaining VLA data with a beam size of $3.1'\times2.8'$. We adopted a second-order polynomials to fit the baselines. The processed VLA data achieved a sensitivity of $5\sigma = 1.4  \times 10^{19}\,{\rm cm^{-2}}$.
 Subsequently, we resampled the VLA data according to the pixel size of FAST and calculated the zeroth moment image of the VLA data within the integrated velocity range utilized by FAST. In Figure\,\ref{fig:compare-NGC2683}, we present the \HI\ flux profiles for both the FAST and VLA observations perpendicular to the plane. We find that, in comparison to the VLA observations, the FAST data reveals higher \HI\ fluxes at large galactic heights, with the enhanced detections extending to positions further away from the major axis. In contrast, the VLA detections are primarily concentrated within $\sim 2.5'$ of the major axis. Specifically, for the flux profiles at offsets of $\pm 8$ arcminutes from the galaxy center (top two panels of Figure\,\ref{fig:compare-NGC2683}), the FAST observational data exhibits a more extended profile compared to the VLA observations. By subtracting the reprocessed VLA data from the FAST data and applying a $3 \times 3$ box smoothing, we also present in Figure\,\ref{fig:compare-NGC2683} the column density maps of the excess \HI\ detected by FAST in comparison to VLA. The column density of the excess \HI\ map is predominantly distributed around $\sim 3 \times 10^{19}\,{\rm cm^{-2}}$, while the few negative regions are likely attributable to pointing uncertainties between the observations, as evidenced by the slightly offset peak \HI\ positions detected by FAST and VLA, as shown in Figure\,\ref{fig:moment0-optical-NGC2683}. The excess \HI\ map clearly demonstrates that, in contrast to the limited sensitivity and missing short spacings of the VLA, FAST has detected a more extended and diffuse \HI\ distribution. In comparison, there is no significant excess in the center of the galaxy, which may be related to the relatively active star formation and higher gas column density in that region. Therefore, FAST does not detect more diffuse gas in the galaxy center than the VLA. We also examined the velocity field (moment 1) of the FAST data after subtracting the VLA residuals, and found that the excess \HI\ exhibits a velocity field that is generally similar to that of NGC~2683.


The galaxy dw1 is classified as a dwarf irregular (dIrr) galaxy and is believed to be a member of the NGC~2683 galaxy group according to \citet{karachentsev2015new}. Based on its $M_{\rm B}$ and $D_{\rm B26}$ values, dw1 is identified as one of the faintest and smallest members within the group \citep{saponara2020new}. It is located to the southeast of NGC~2683 at a projected distance of approximately $7'$. The distance of dw1 remains undetermined, so we employ the distance of NGC~2683 as a substitute. As shown in left panel of Figure\,\ref{fig:moment0-optical-NGC2683}, our observations using the FAST reveal that dw1 is situated adjacent to the extended disk of NGC~2683. The center of \HI\ emission in dw1 deviates from its optical center and appears to be skewed towards the NGC~2683 disk. \citet{saponara2020new} presented VLA observations of dw1 and also reported a departure between the \HI\ and optical emissions. They achieved a detection limit of $5\sigma = 4.4\times10^{19}\,\rm{cm^{-2}}$, which is significantly higher than our FAST observations with a detection limit of $5\sigma = 1.5\times10^{18}\,\rm{cm^{-2}}$. The total \HI\ mass of dw1 is determined to be $M_{\rm HI} = 0.2 \times 10^7\,{\rm M_\odot}$. Figure\,\ref{fig:moment1-NGC2683} present the \HI\ velocity field of NGC~2683 and its satellite galaxies. The velocity field of NGC~2683 show an increasing from northeast to southwest, while the satellite galaxy dw1 has no obvious velocity gradients.

\begin{figure*}
\centering
	\includegraphics[width=0.95\textwidth]{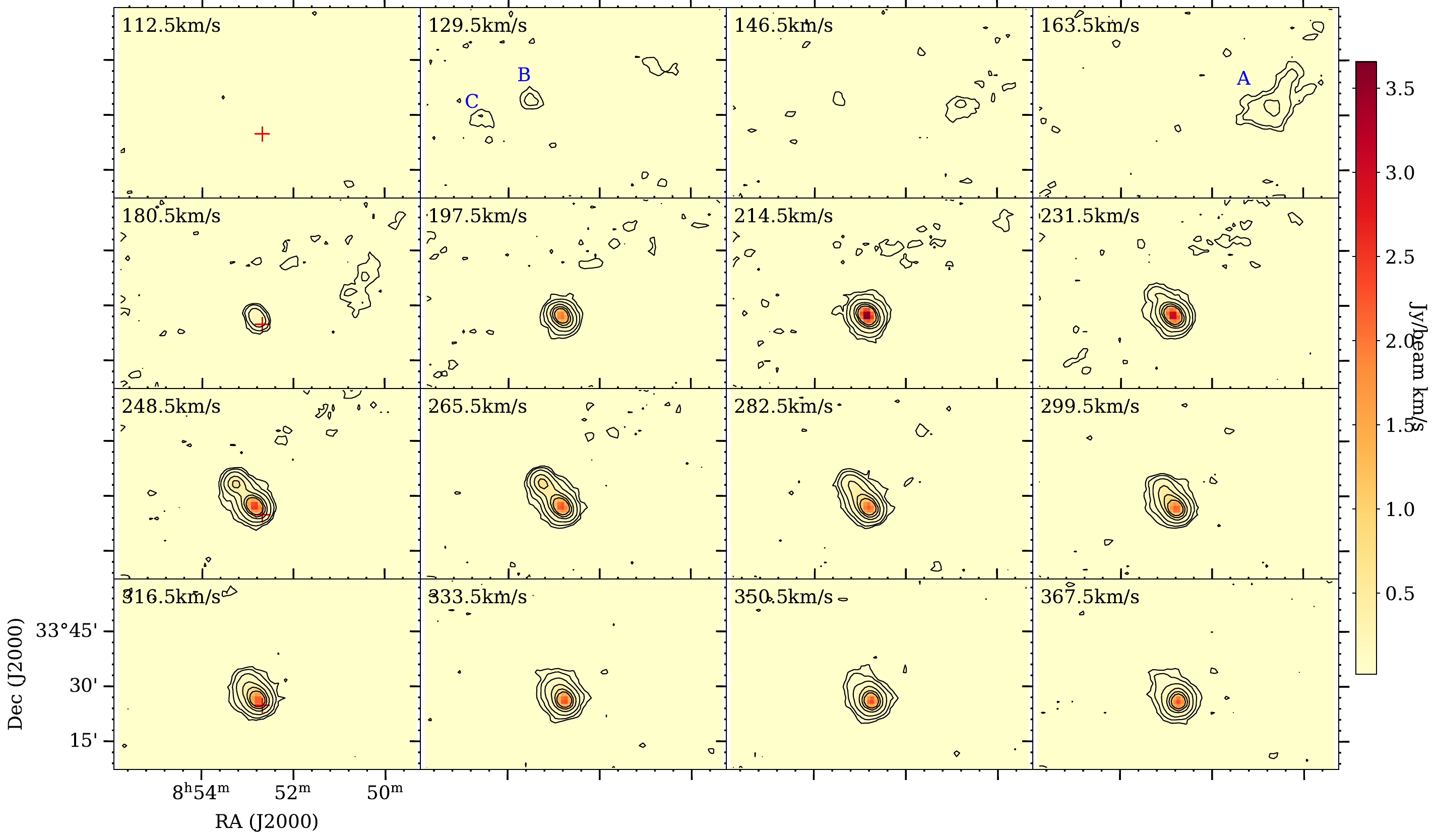}
	\includegraphics[width=0.95\textwidth]{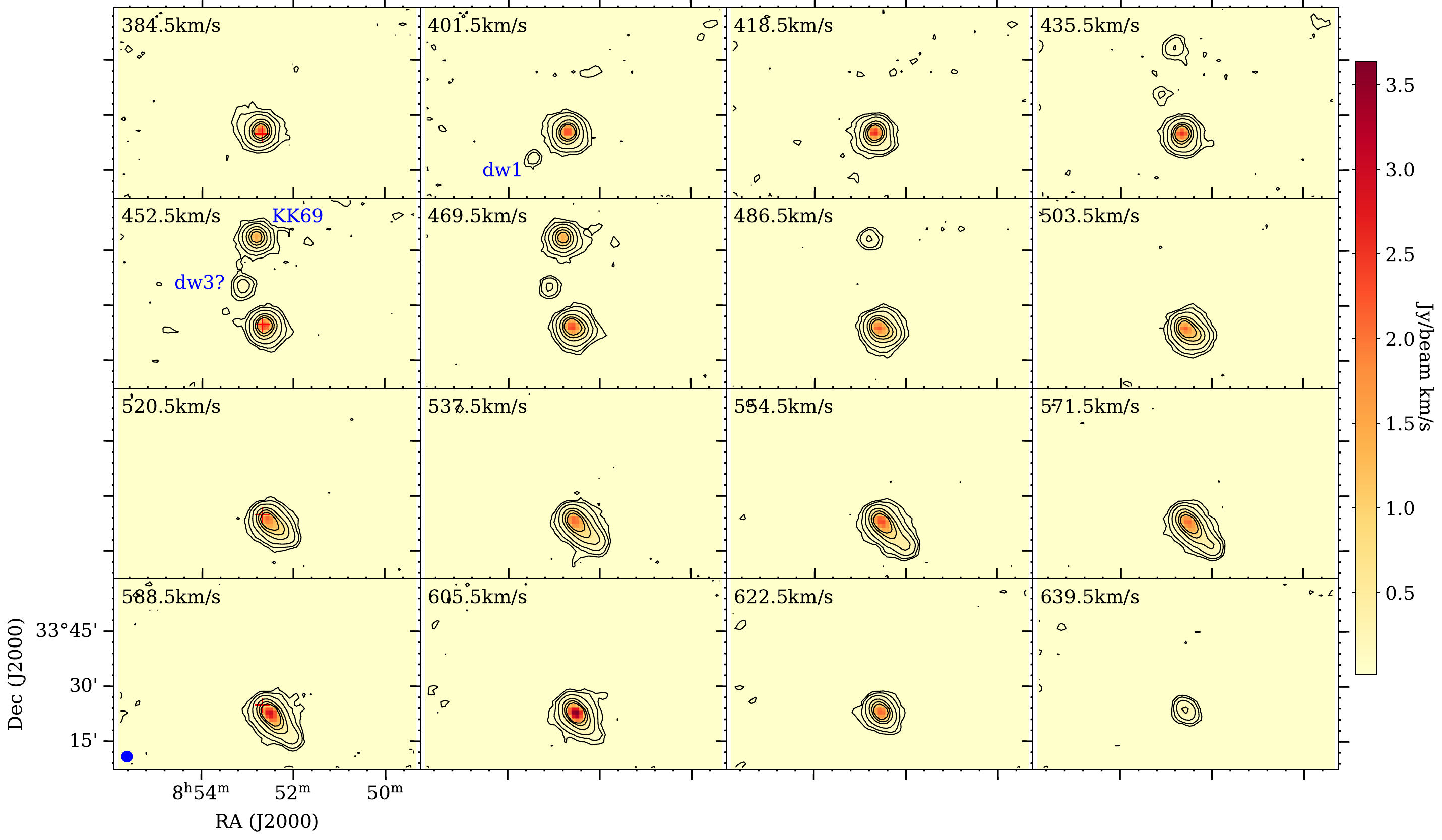}
	\caption{The \HI\ channel map of the NGC~2683 group obtained from the FAST observations. Each channel corresponds to a velocity range of 17 $\rm{km\,s^{-1}}$. The contour levels are set at 1, 2, 5, 15, 30, and 50 times the 5$\sigma$ level, where $\sigma$ represents the noise level of $0.5\,\rm{mJy/beam}$ for each channel map. The center of NGC~2683 is indicated by a red plus symbol. The blue filled circle indicates the beam size of $2.9'$ of FAST.}

	\label{fig:Channel-maps}
\end{figure*}

In the right panel of Figure\,\ref{fig:moment0-optical-NGC2683}, we present the \HI\ integrated flux density contours obtained from FAST, GMRT, and VLA observations overlaid on the DECaLs optical image for the galaxies KK~69 and dw3$?$. Our FAST observations achieved a detection limit of $5\sigma = 1.5\times10^{18}\,\rm{cm^{-2}}$, which is significantly lower than the GMRT observations of KK~69 with $5\sigma = 1.5\times10^{19}\,\rm{cm^{-2}}$ and the VLA observations of dw3$?$ with $5\sigma = 4.4\times10^{19}\,\rm{cm^{-2}}$ as reported by \citet{saponara2020new}. Notably, both the peak \HI\ emission from the FAST and GMRT observations do not coincide with the center of the stellar distribution in KK~69. Our FAST observations reveal an extended \HI\ distribution with a tail extending towards dw3$?$ and NGC~2683 in the southeast direction of KK~69. This feature is also evident in the channel maps at centre velocity of $\rm{452.5\,km\,s^{-1}}$ shown in Figure\,\ref{fig:Channel-maps}, where the tail of KK~69 appears to be almost adjacent to dw3$?$ region. The nature of dw3$?$ is discussed as an \HI\ cloud associated with the host galaxy NGC~2683, as no optical counterpart has been observed \citep{crosby2023new}. As illustrated in Figure\,\ref{fig:moment1-NGC2683}, the velocity range of dw3$?$ closely resembles that of KK~69, and the velocity of the connecting tail between KK~69 and dw3$?$ is akin to the northern connection part. These observations imply a potential interaction between KK~69 and NGC~2683, with dw3$?$ possibly being a product of this interaction and potentially originating from KK~69. The distance of dw3$?$ is ambiguous, thus we adopt the distance of KK~69 as a substitute. The total \HI\ mass of KK~69 is calculated to be $M_{\rm HI} = 9.6 \times 10^7\,{\rm M_\odot}$, which is more than double the value of $M_{\rm HI} = 4.2 \times 10^7\,{\rm M_\odot}$  reported in \citet{saponara2020new}. As for dw3$?$, our estimate \HI\ mass is $M_{\rm HI} = 0.8 \times 10^7\,{\rm M_\odot}$.

The \HI\ flux of complex B and C is 0.07 and 0.09 $\rm{Jy\,km\,s^{-1}}$ respectively, with a sensitivity of $5\sigma = 0.03\,\rm{Jy/beam\,km\,s^{-1}}$, corresponding to a column density sensitivity of $5\sigma = 1.1\times10^{18}\,\rm{cm^{-2}}$. For complex A, the \HI\ flux is 1.4\,$\rm{Jy\,km\,s^{-1}}$ with a sensitivity of $5\sigma = 0.04\,\rm{Jy/beam\,km\,s^{-1}}$, corresponding to a column density sensitivity of $5\sigma = 1.4\times10^{18}\,\rm{cm^{-2}}$. The discussion of these three complexes can be found in the subsequent section.

\begin{figure*}
	\includegraphics[width=0.95\textwidth]{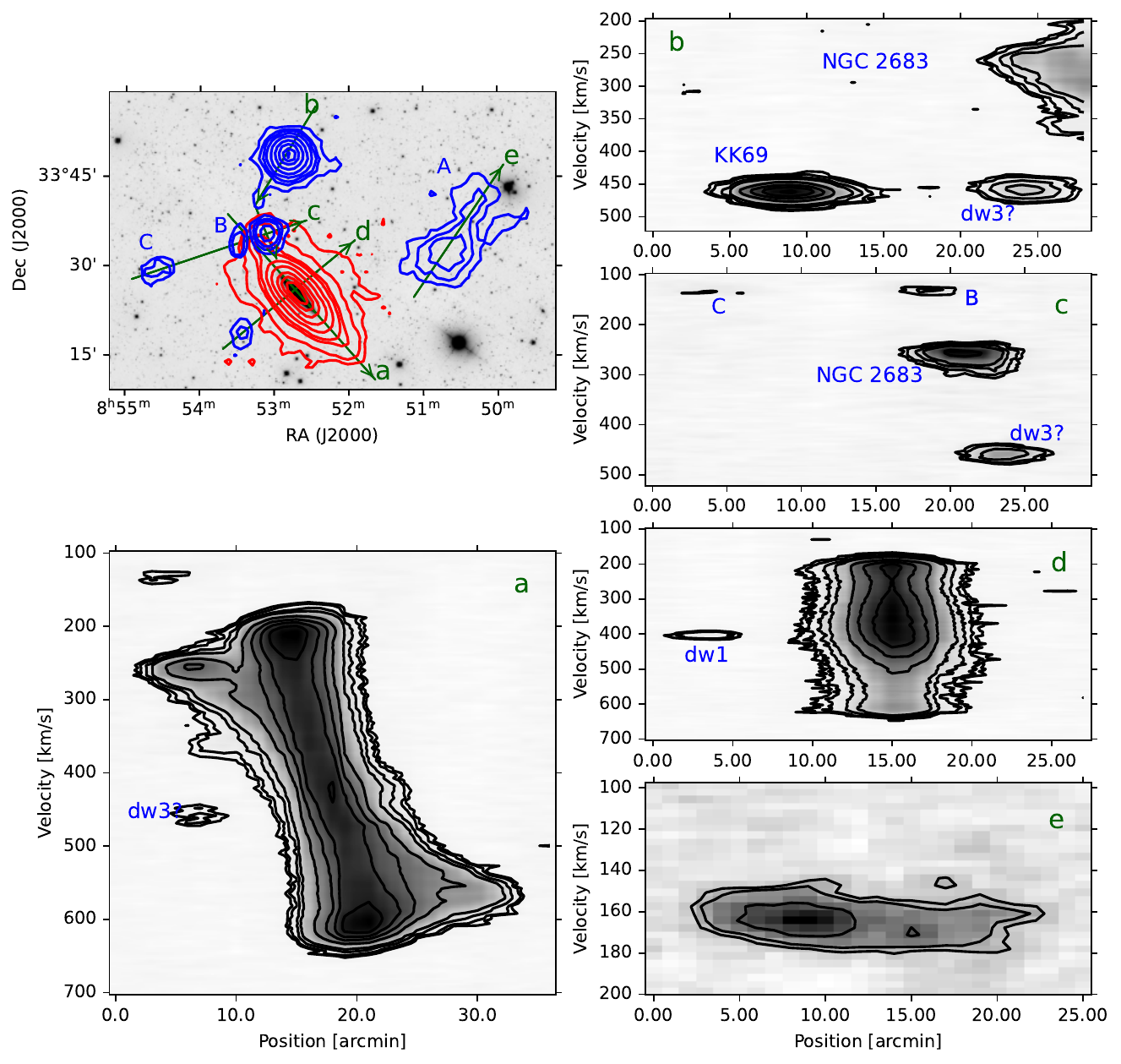}
    \caption{We present five PV plots obtained along different directions. The top left panel corresponds to the same image as Figure\,\ref{fig:moment0-optical}, and the positions and orientations of the PV diagrams in panels a $-$ e are indicated by dark green arrows. The contour levels are set at 1, 1.5, 3, 5, 15, 30, 50, 80, and 120 times the 5$\sigma$ level (except for panel b which times 3$\sigma$ level), where $\sigma$ is $0.4\,\rm{mJy/beam}$ for panel c and $0.5\,\rm{mJy/beam}$ for panels a, b, d, and e. These implies \HI\ column density sensitivities of $5\sigma = 2.5\times10^{17}\,\rm{cm^{-2}}$ and $5\sigma = 3.1\times10^{17}\,\rm{cm^{-2}}$ over velocity resolution of 3.4 $\rm{km\,s^{-1}}$, respectively.}
    \label{fig:PVMaps}
\end{figure*}

Figure\,\ref{fig:Channel-maps} displays the channel maps of the NGC~2683 group, revealing the presence of complexes A, B, and C in the low-velocity channels. These regions show limited overlap in velocity range with NGC~2683, except for complex A, where the low-velocity part of NGC~2683 and the high-velocity part of complex A both appear at a center velocity of $\rm{180.5\,km\,s^{-1}}$ in Figure\,\ref{fig:Channel-maps}. 
Figure\,\ref{fig:PVMaps} illustrates the Position-Velocity (PV) diagram of the NGC~2683 group. Panel b of Figure\,\ref{fig:PVMaps} indicates that the velocity of dw3$?$ is closer to the velocity of KK~69 compared to the nearest part of NGC~2683, and panel c also indicates that the velocities of complexes B and C are both smaller than that of NGC~2683.

\section{Discussion}

\begin{table*}
	\centering
	\caption{The basic information of complexes A, B, and C.}
	\label{table:complexs}
	\begin{tabular}{lccccccccr} 
		\hline
		Name &  Central velocity & Velocity FWHM & peak column density & a $\times$ b  & $M_{\rm HI}$ & $M_{\rm dyn}$ & $\bar{n}_{\rm HI}$ \\
		    & $\rm{km\,s^{-1}}$ & $\rm{km\,s^{-1}}$ &  $10^{18}\,\rm{cm^{-2}}$ & $'$ & ${\rm\times d^2\,[10^5\,M_\odot]}$ & ${\rm\times d\, [10^7\,M_\odot]}$ & ${\rm \times d^{-1}\, [10^{-3}\, cm^{-3}]}$ \\
		\hline
		A & 163.5 & 24.2 & 5.8 & $20'\times 8'$ & 3.36 & 4.59 & 0.52 \\
		B & 132.1 & 24.5 & 2.2 & $5'\times 3'$  & 0.16 & 1.44 & 0.85 \\
		C & 131.6 & 22.3 & 2.0 & $5'\times 4'$  & 0.20 & 1.38 & 0.72 \\
		\hline
	\end{tabular}
\end{table*}

\begin{figure*}
	\includegraphics[width=0.32\textwidth]{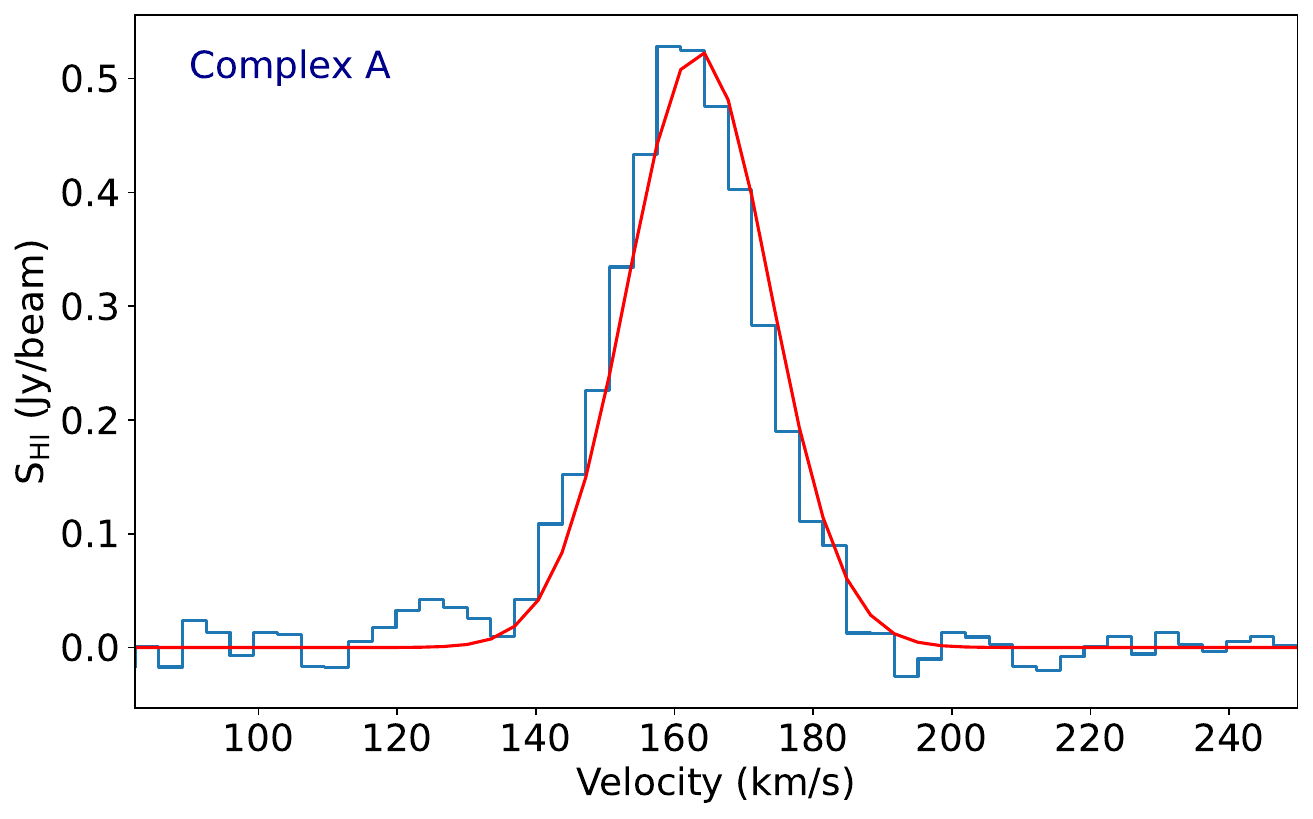}
	\includegraphics[width=0.33\textwidth]{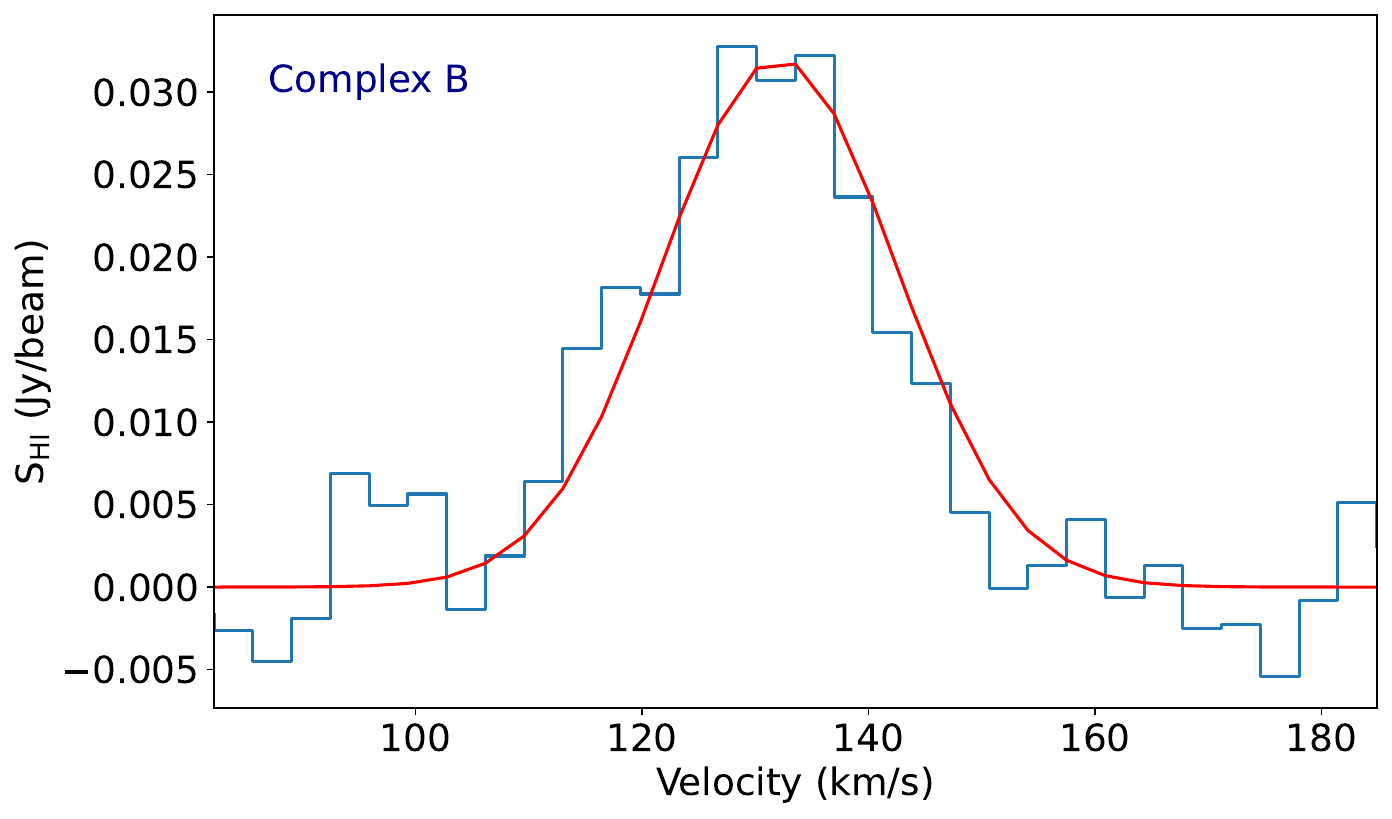}
	\includegraphics[width=0.33\textwidth]{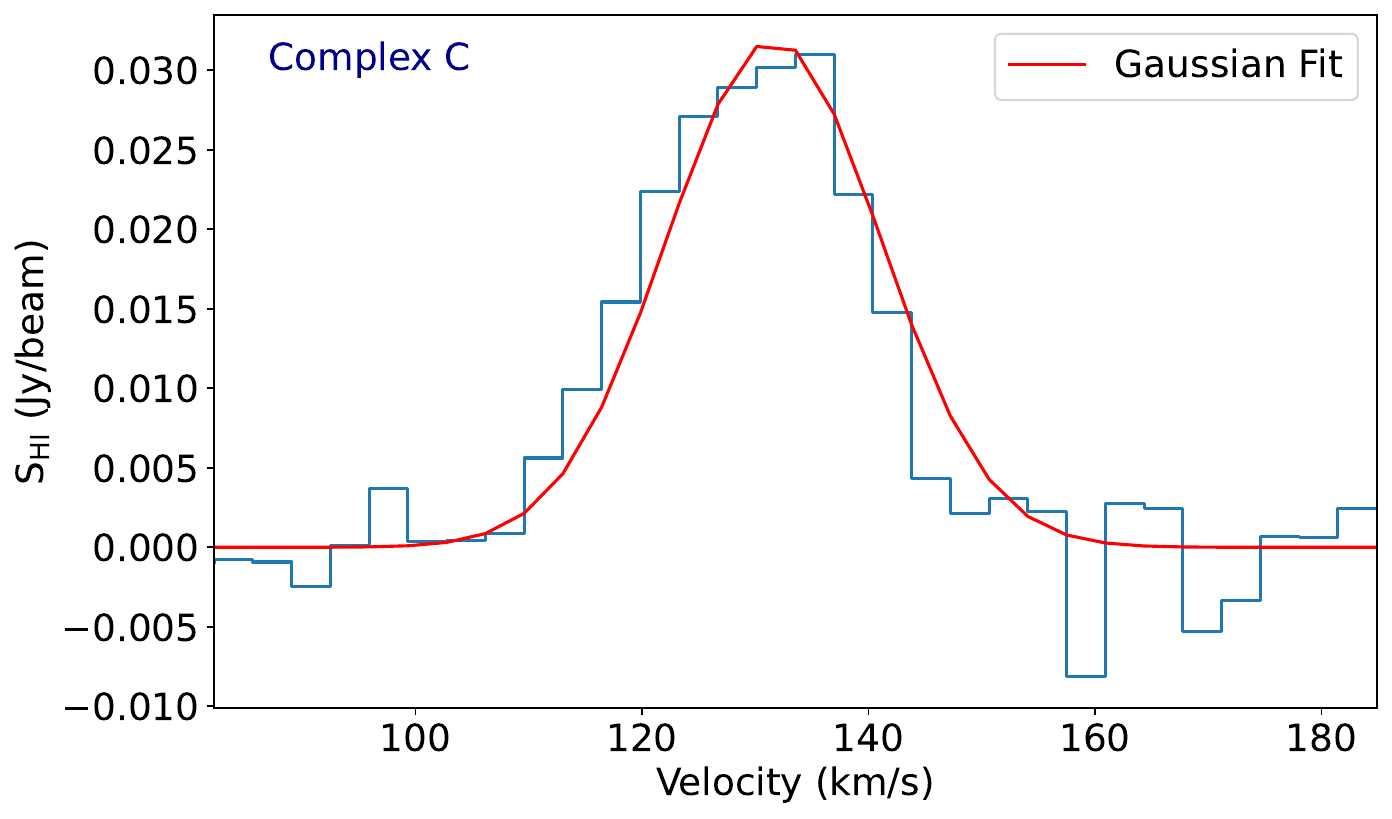}

    \caption{The integrated \HI\ spectra of complexes A, B, and C, obtained from the FAST observations, are presented in the left, middle, and right panels, respectively. The red curves overlaid on the spectra represent the results of Gaussian fitting.}
    \label{fig:complex_spectrum}
\end{figure*}

\subsection{Three HI complexes}

We have identified three anomalous velocity clouds labeled as complexes A, B, and C. It is hypothesized that these complex clouds observed in external galaxies could be analogous to the intermediate- and high-velocity clouds (IVCs and HVCs) found in the Milky Way \citep{wakker1997high}. Figure\,\ref{fig:complex_spectrum} presents the integrated \HI\ spectra of complexes A, B, and C. The central velocities and velocity full widths at half-maximum (FWHM) of the clouds were derived through Gaussian fitting, and the corresponding results are tabulated in Table\,\ref{table:complexs}. We use ellipses to approximate the sizes of the clouds, with the major and minor diameters of the three complexes listed in Table\,\ref{table:complexs}. Furthermore, Table\,\ref{table:complexs} also provides the peak \HI\ column densities, neutral hydrogen masses of $M_{\rm HI}$, indicative dynamical masses within the \HI\ extent of $M_{\rm dyn}$, and mean neutral hydrogen number densities of $\bar{n}_{\rm HI}$ for each complex. The dynamical masses are calculated using the formula $M_{\rm dyn}=6.2 \times 10^3\,\bar{a}\, W_{50}^2\, d$, where $\bar{a}=\sqrt{ab}$ represents the average angular diameter, $W_{50}$ is the velocity FWHM, and the distance $d$ is in Mpc. The mean atomic number density is given by $\bar{n}_{\rm HI} = 0.74\,S\, \bar{a}^{-3}\, d^{-1}\,{\rm cm^{-3}}$, with $S$ denoting the integrated \HI\ flux density in units of $\rm{Jy\,km\,s^{-1}}$.

\citet{adams2013catalog} presented a catalog comprising 59 ultra-compact high-velocity clouds (UCHVCs), extracted from the 40\% complete ALFALFA \HI-line survey. These ALFALFA UCHVCs exhibit median flux densities of 1.34$\,\rm{Jy\,km\,s^{-1}}$, median angular diameters of 10$'$, and median velocity widths of 23$\,\rm{km\,s^{-1}}$. Complexes A, B, and C exhibit similar line widths and sizes to the UCHVCs. Although complex B overlaps with NGC~2683 in projection, it demonstrates a velocity deviation of approximately 120$\,\rm{km\,s^{-1}}$ from the velocity of NGC~2683 at the same spatial location. The significant velocity differences observed in complexes A, B, and C relative to NGC~2683 suggest the possibility that they could  be UCHVCs within the local group. Determining the distances of high-velocity clouds (HVCs) in the Milky Way remains a challenge. Assuming a typical distance of $\sim 1\,$Mpc for these complexes, the estimated \HI\ masses of A, B, and C would be $3.36\times 10^5\,{\rm M_\odot}$, $0.16 \times 10^5\,{\rm M_\odot}$, and $0.20 \times 10^5\,{\rm M_\odot}$, respectively. The major and minor diameters of these complexes are $5.8\times 2.3\,{\rm kpc}$, $1.5\times 0.9\,{\rm kpc}$, and $1.5\times 1.2\,{\rm kpc}$, with indicative dynamical masses of $4.59\times 10^7\,{\rm M_\odot}$, $1.44\times 10^7\,{\rm M_\odot}$, and $1.38\times 10^7\,{\rm M_\odot}$, respectively. In comparison with the findings from \citet{adams2013catalog}, complexes B and C are also considered relatively compact in nature. However, their flux densities are lower than those reported for UCHVCs. Given that \citet{adams2013catalog} identified only 59 UCHVCs across nearly 2800 square degrees, the likelihood of three UCHVCs occurring in the NGC~2683 group region is low. Therefore, complexes B and C are more likely to be HVCs within the NGC~2683 group, while complex A is likely an HVC within the Milky Way.

Assuming that complexes A, B, and C are high-velocity clouds (HVCs) within the NGC~2683 group and at the same distance as NGC~2683, the major and minor diameters of these complexes would be $54.5\times 21.8\,{\rm kpc}$, $13.6\times 8.2\,{\rm kpc}$, and $13.6\times 10.9\,{\rm kpc}$, respectively. The estimated \HI\ masses of complexes A, B, and C are $2.94 \times 10^7\,{\rm M_\odot}$, $0.14 \times 10^7\,{\rm M_\odot}$, and $0.18 \times 10^7\,{\rm M_\odot}$, with indicative dynamical masses of $4.30\times 10^8\,{\rm M_\odot}$, $1.35\times 10^8\,{\rm M_\odot}$, and $1.29\times 10^8\,{\rm M_\odot}$, respectively. Complexes A and C are located approximately $15'$ northwest and northeast of the plane of NGC~2683 in projection, corresponding to a distance of approximately 40$\,$kpc.

\begin{figure*}
\centering

	\includegraphics[width=0.9\textwidth]{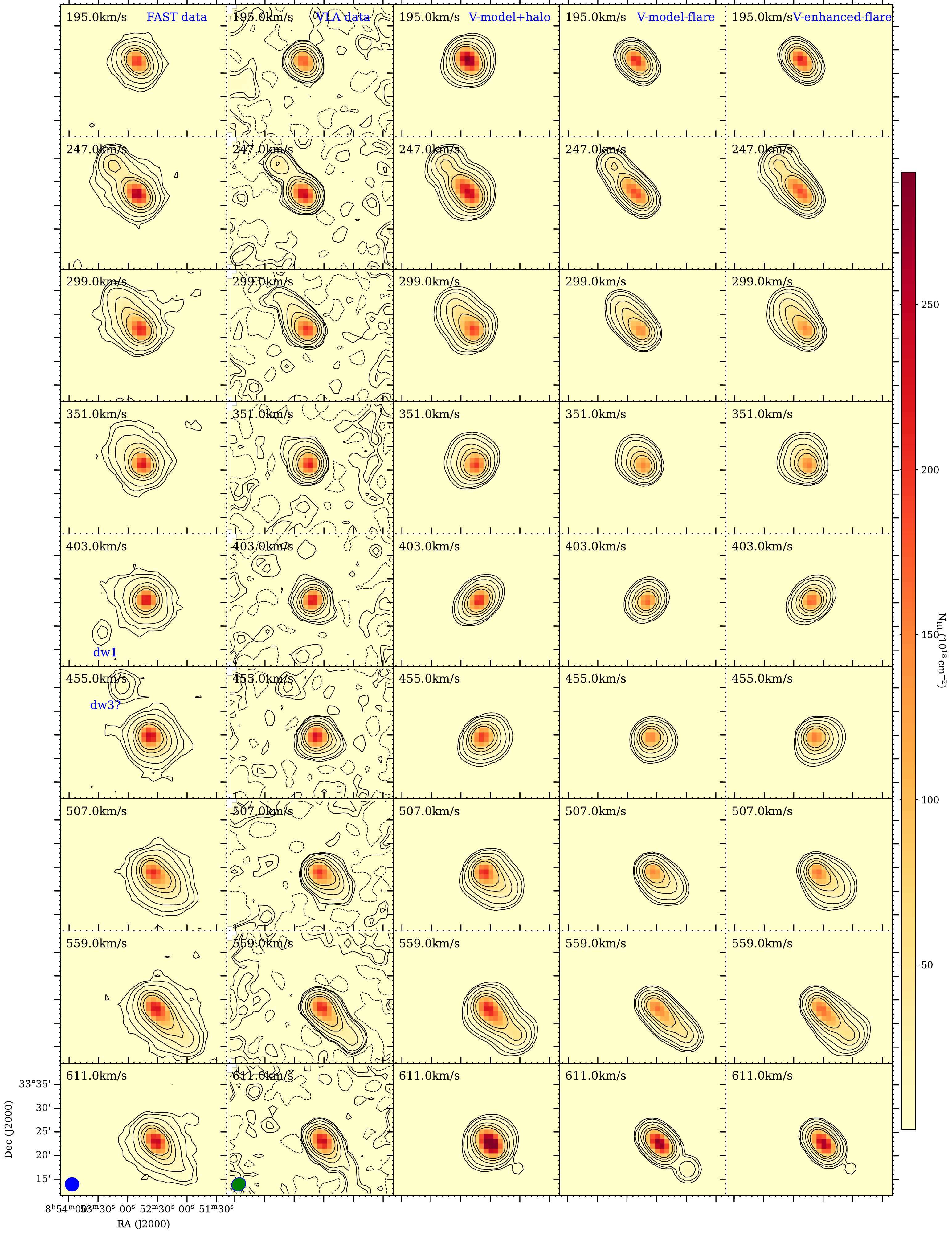}

	\caption{The panels in the first column represent the channel maps from FAST, while the second column displays the channel maps of reprocessed VLA data. The third column illustrates the channel maps derived from the foundational model proposed by \citet{vollmer2015flaring}, which includes a halo component (designated as V-model+halo). The fourth column presents the channel maps considering the flare model (referred to as V-model-flare) from \citet{vollmer2015flaring} . The fifth column represents the V-model+halo, with the halo component removed for $R < 14.6\,$kpc (termed V-enhanced-flare). All three models have been smoothed to the resolution of FAST. Each channel corresponds to a velocity range of 52 $\rm{km\,s^{-1}}$. The contour levels are defined at 1, 2, 5, 15, 30, and 50 times the $5\sigma = 1.2\times10^{18}\,\rm{cm^{-2}}$, where $\sigma$ denotes the noise level of $0.5\,\rm{mJy/beam}$ observed by FAST. The dashed line in the second column represents the $-5\sigma$ contour.
	The blue filled circle indicates the beam size of $2.9'$ for FAST, while the green filled circle denotes the VLA beam size of $3.1'\times2.8'$. }
	\label{fig:Channel-maps-compare}
\end{figure*}

\subsection{\HI\ halo}

\begin{figure*}
\centering
	\includegraphics[width=0.5\textwidth]{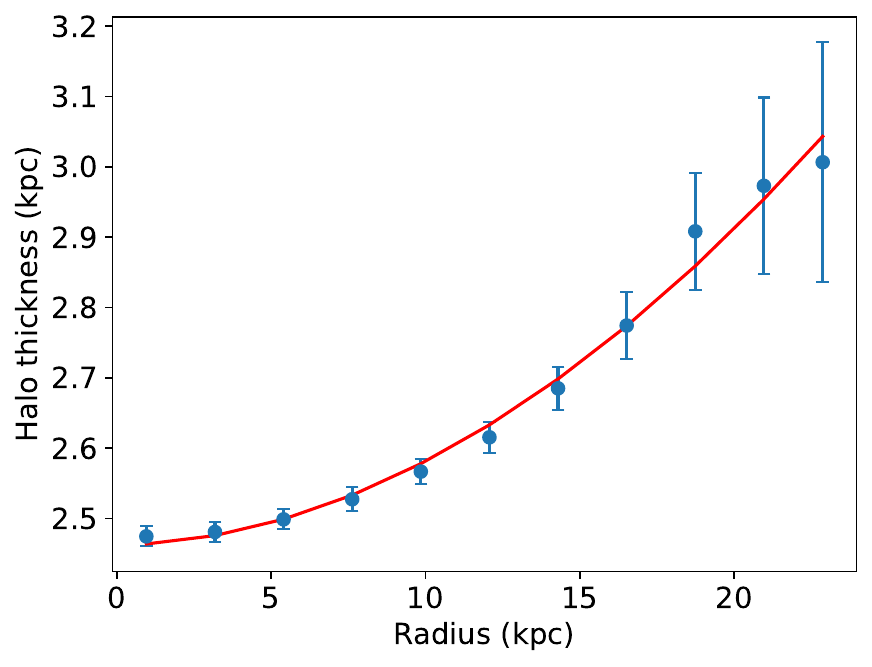}

	\caption{Gas distribution in the region of the halo. Scale height of the gaseous halo $z$ (circles) with a power-law fit (solid red line).}
	\label{fig:scale_height}
\end{figure*}

\begin{figure*}
\centering
	\includegraphics[width=0.33\textwidth]{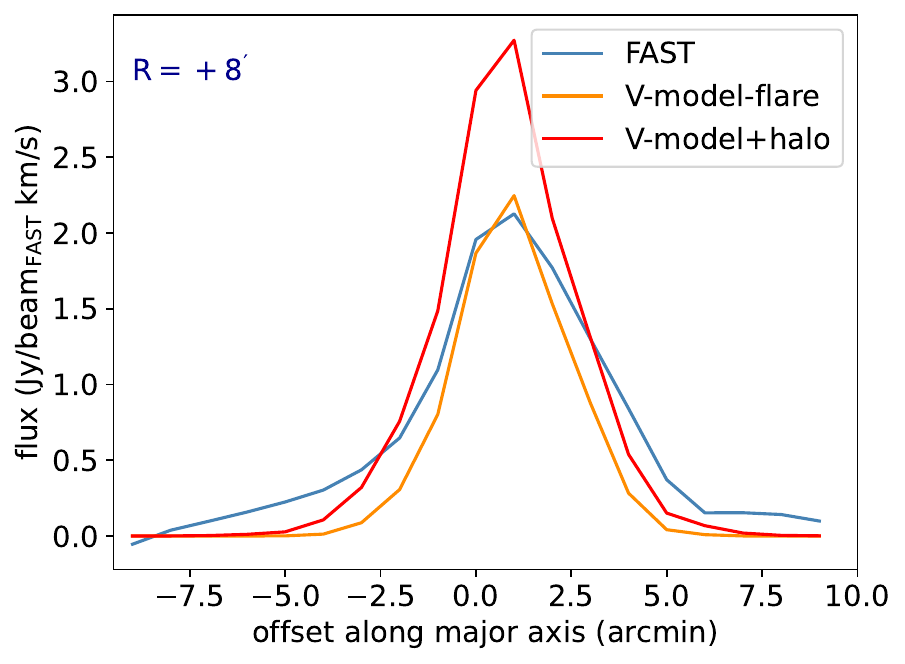}
	\includegraphics[width=0.33\textwidth]{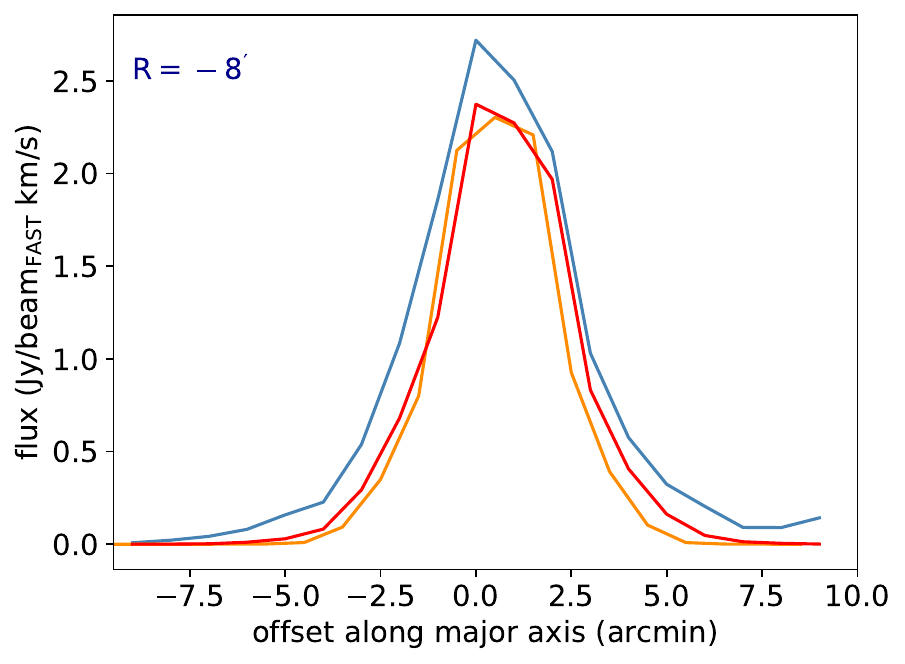}\\
		\includegraphics[width=0.33\textwidth]{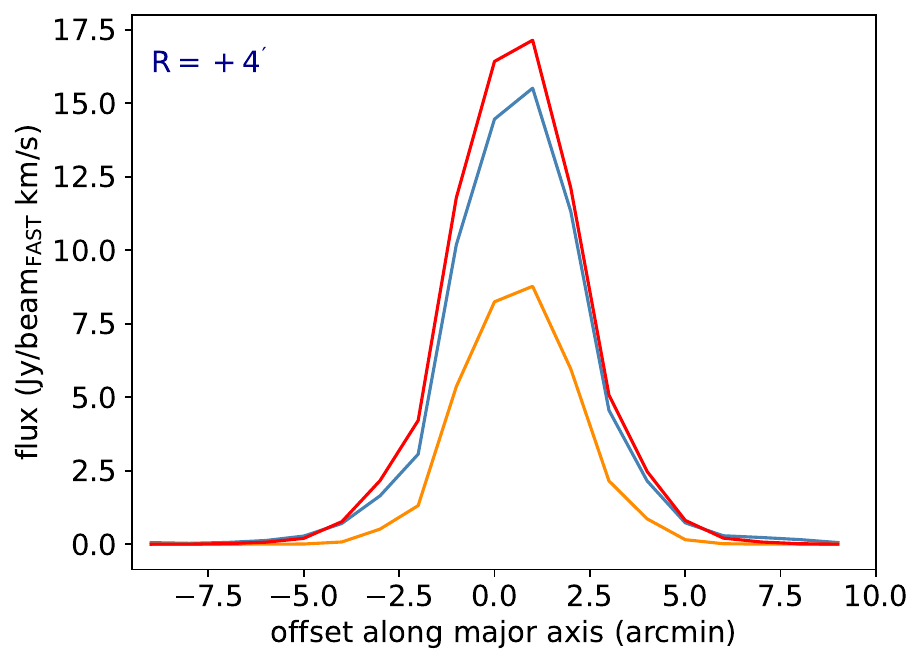}
	\includegraphics[width=0.33\textwidth]{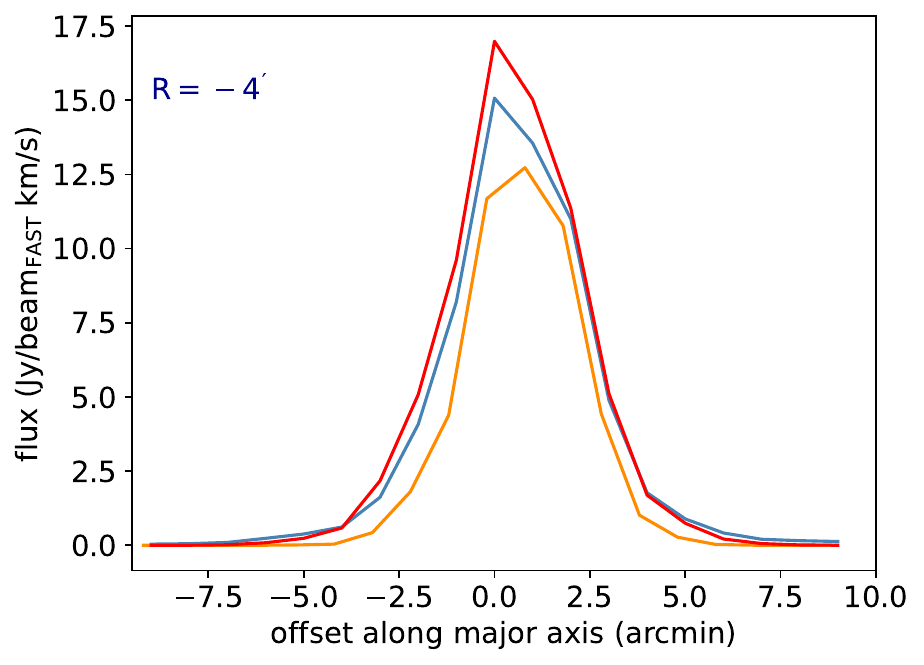}\\
	\includegraphics[width=0.33\textwidth]{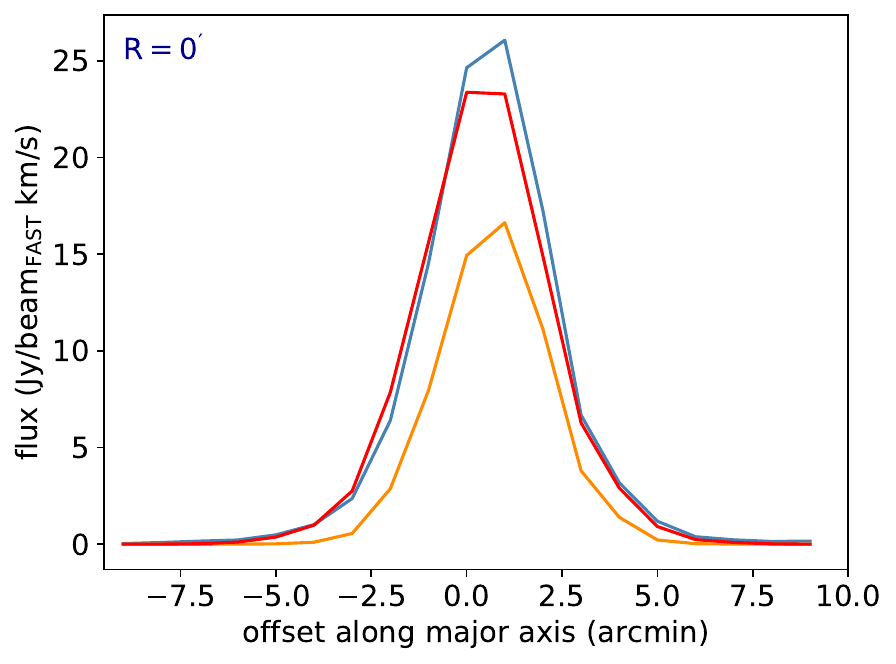}
	\includegraphics[width=0.34\textwidth]{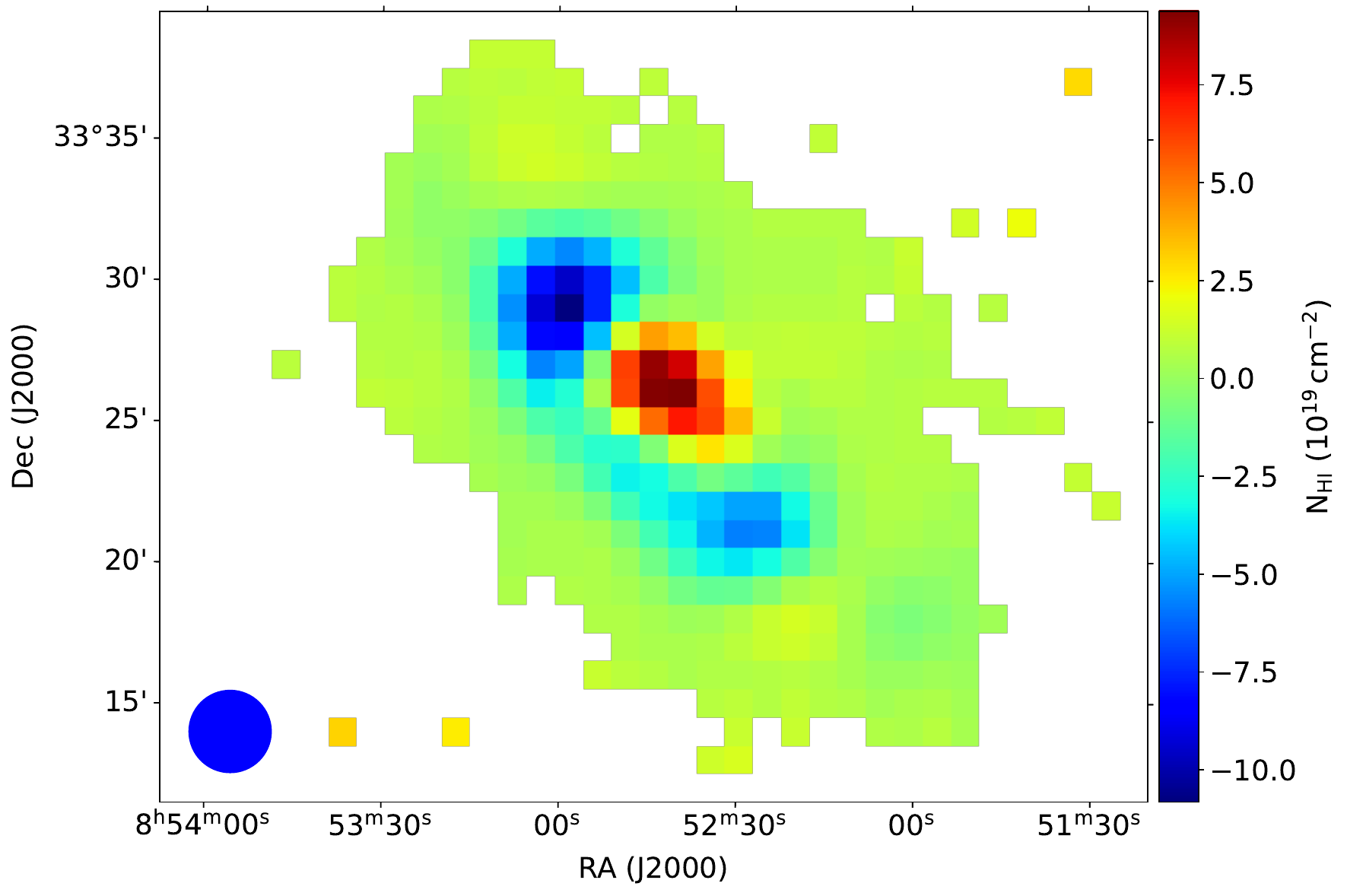}

    \caption{The left panels and the top two panels on the right respectively illustrate the flux profiles perpendicular to the galactic plane at different distances from the galactic center along the major axis. R in each panel denotes the distance from the center, with positive offsets directed towards the north-east. The steel-blue lines represent the FAST observationals, while the dark orange lines illustrate the profiles of the V-model-flare model. The red lines depict the profiles of the V-model+halo model. The bottom right panel presents the column density maps of the excess \HI\ detected by FAST, compared to the V-model+halo model [FAST$-$(V-model+halo)]. The solid blue circle indicate the beam size of FAST.}
    \label{fig:compare-FAST_Vollmer_NGC2683}
\end{figure*}

\begin{figure*}
\centering

	\includegraphics[width=0.9\textwidth]{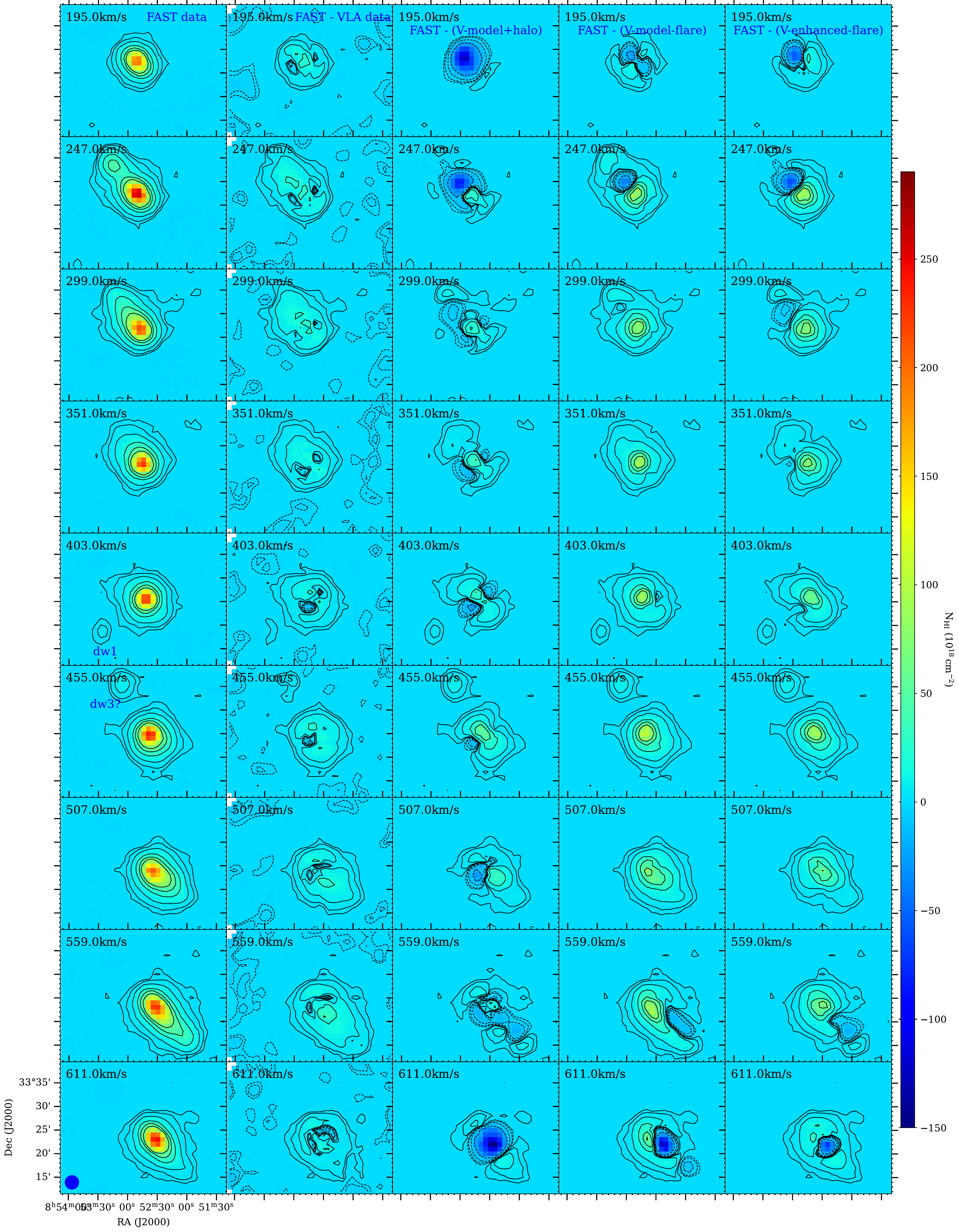}

	\caption{Similar to Figure\,\ref{fig:Channel-maps-compare}, but this figure showcases FAST data along with its residual channel maps in comparison to VLA data, as well as the models V-model+halo, V-model-flare, and V-enhanced-flare, respectively. All model data are presented at the resolution of FAST. The solid contour lines are defined at 1, 2, 5, 15, 30, and 50 times the $5\sigma = 1.2\times10^{18}\,\rm{cm^{-2}}$, while the dashed contour lines represent -15, -5, -2, and -1 times the $5\sigma = 1.2\times10^{18}\,\rm{cm^{-2}}$, where $\sigma$ denotes the noise level of $0.5,\rm{mJy/beam}$ observed by FAST. 
	The blue filled circle indicates the beam size of $2.9'$ for FAST.}
	\label{fig:residual-Channel-maps-compare}
\end{figure*}

The neutral gas composition of a spiral galaxy can be broadly categorized into three distinct components: a compact molecular disk, an atomic gas disk, and an extended atomic gas halo. While the internal gas disk is exceptionally thin (e.g., $\sim$\,100\,pc for the Milky Way, \citealt{kalberla2009hi}), the potential gas halo component extends significantly further, ranging in size from a few kiloparsecs, as observed in NGC~2403 \citep{fraternali2002deep} and NGC~3198 \citep{gentile2013halogas}, up to approximately 14 kiloparsecs in the case of NGC~891 with a filament extending up to 22 kpc \citep{oosterloo2007cold}. The fraction of \HI\ gas in the halo of spiral galaxies appears to vary significantly among different galaxies \citep{sancisi2008cold, putman2012gaseous}, with values ranging from approximately 10\% found in NGC~2403 \citep{fraternali2002deep} to 30\% in NGC~891 \citep{oosterloo2007cold}.

Using the \HI\ data of NGC~2683 observed with VLA, \citet{vollmer2015flaring} made different 3D models to produce model \HI\ data cubes, and conducted a comparative analysis with observed data in order to investigate the 3D structure of the atomic gas disk. Their models consist primarily of the following components: (i) a thin gas disk with a thickness of 500 pc; (ii) varying gas flares beyond galactic radii of 9 kpc; (iii) a possible warp of the disk; and (iv) an outer gas ring ($R > 25\,$kpc). They deliberated that the vertical expansion of \HI\ results from the projection of a flaring gas disk, and ruled out the presence of an extensive atomic gas halo surrounding the optical and thin gas disk. Nonetheless, our high-sensitivity observational data suggest a high likelihood of the presence of a gas halo.

As shown in Figure\,\ref{fig:moment0-optical-NGC2683}, the \HI\ observations from VLA in \citet{vollmer2015flaring} reveal that atomic hydrogen is distributed over a diameter of $26.5'$, nearly three times the optical diameter. However, there is limited extension in the vertical direction, approximately double the optical range.  Our observations with FAST indicate a slightly larger diameter compared to VLA.  Particularly noteworthy is the vertical extent of our \HI\ distribution, which is approximately four times that of the VLA observations (with FAST detections around 18.7$'$ compared to approximately 4.5$'$ for VLA). Nevertheless, the flare model proposed by \citet{vollmer2015flaring} encounters challenges in replicating such an extensive distribution of \HI.

 \citet{vollmer2015flaring} presented model channel maps in their Figure\,C.4. and Figure\,C.5, featuring components such as a thin disk, an elliptical component, an inclined warp, a radially decreasing velocity dispersion, a flare, and an \HI\ halo. By comparing these model channel maps with VLA observations, they found that the \HI\  distribution in the model incorporating an \HI\ halo extended further than what was observed by the VLA, leading them to rule out the possibility of the existence of an \HI\ halo with a column density in excess of $\sim 3 \times 10^{19}\,{\rm cm^{-2}} $. This threshold value also corresponds to the majority of the residual values observed by FAST that exceed those detected by VLA, as shown in Figure\,\ref{fig:compare-NGC2683}. In Figure\,\ref{fig:Channel-maps-compare},  we present the channel maps obtained from both FAST and VLA observations. The contour levels are defined starting from the $5\sigma$ of the FAST observation. It can be observed that, at the resolution similar to FAST, the detection region of VLA remains significantly smaller than that of FAST. This discrepancy can be attributed to the VLA interferometer's lack of short-spacing information and insufficient sensitivity, which results in the inability to capture the extended structures observed by FAST. Additionally, we convolved the best-fit model from \citet{vollmer2015flaring}, specifically the flare F3 model (hereafter referred to as the V-model-flare), to the beam size of FAST and displayed its channel maps in Figure\,\ref{fig:Channel-maps-compare}. While the V-model-flare model demonstrates a reasonable alignment with the VLA data, each channel map reveals a detection range that is notably smaller than that of FAST. This suggests that the V-model-flare, which solely considers the flare component, is inadequate for explaining the high sensitivity observations made by FAST.

To address this, we utilized the model of V-model-flare from \citet{vollmer2015flaring}, incorporating an \HI\ halo component, and computed the channel maps of this modified model at FAST's resolution, as shown in Figure\,\ref{fig:Channel-maps-compare}. For the \HI\ halo, we referenced to Equation 8 from \citet{vollmer2015flaring}, which is expressed as $\rho_{\rm halo} = \chi \, \rho_{\rm disk}\, \frac{{\rm sinh}(z/z_0)}{{\rm cosh}(z/z_0)^2}$, where $\rho_{\rm disk}$ represents the density of the gas disk, and $z_0$ is the vertical scale height  in kpc. This is an empirical function that represents well the vertical halo gas distribution in NGC~891 \citep{oosterloo2007cold}, and also describes a sample of 15 nearby late-type galaxies from the Hydrogen Accretion in LOcal GAlaxieS (HALOGAS) survey \citep{marasco2019halogas}. Typically, the \HI\ scale heights are nearly constant in the inner region of the galaxy and increase as a function of radius in the outer region \citep{dickey1990hi, kalberla2009hi, oosterloo2007cold}. However, considering the low spatial resolution of FAST, it is challenging to distinctly resolve the disk and the outer halo, making it difficult to obtain the distribution of scale heights as a function of radius from FAST data. 
Therefore, we utilized VLA data to calculate the scale height and performed a power law fitting to obtain the scale height distribution of the galaxy. We employed the VLA moment 0 data smoothed to the resolution of FAST. Initially, we obtained the flux profile along positions parallel to the minor axis at varying distances from the galactic center along the major axis. We then subtracted the contributions of the disk and flare using the Gaussian profile generated from the disk height values of the flare in the V-model-flare model (with the component smoothed to the resolution of FAST). Subsequently, we applied the aforementioned equation to fit the scale height values. Finally, we used a power law to fit the relationship between the scale height and radius, as shown in Figure\,\ref{fig:scale_height}. The parameter $\chi$ is derived from the comparison of the flux obtained from the model with the FAST observations, for which we adopted $\chi = 0.1$.
As show in Figure\,\ref{fig:Channel-maps-compare}, the channel maps of the model incorporating the \HI\ halo (hereafter referred to as V-model+halo) exhibit a spatial distribution that is significantly more extensive than that of the V-model-flare model based on flares, aligning more closely with the observations from FAST. Consequently, we conclude that such an extended \HI\ distribution is likely attributed to the \HI\ halo rather than to flares.
This highlights the crucial role of high-sensitivity observations from the FAST telescope in advancing our comprehension of the extended \HI\ distribution in galaxies and facilitating the study of galactic evolution.

In the following we further investigated if an \HI\ halo around the thin disk is needed to explain the FAST observations. To do so, the model \HI\ halo was set to zero for galactic radii smaller than $14.6\,$kpc, as \citet{vollmer2015flaring} found that the flare begins only at galactic radii greater than $12\,$kpc (using a distance of $7.7\,$Mpc; this corresponds to $14.6\,$kpc when a distance of $9.36\,$Mpc is applied). In Figure\,\ref{fig:Channel-maps-compare}, we present the channel maps of this model (referred to as V-enhanced-flare) and observe that the increasing flare height results in a more extended distribution along the galaxy's minor axis at large galactic radii. However, for smaller radii within the galaxy, the \HI\ distribution along the minor axis is significantly smaller than that observed by FAST. Therefore, we conclude that an \HI\ halo around the thin disk is needed to reproduce the FAST observations.


In Figure\,\ref{fig:compare-FAST_Vollmer_NGC2683}, we present the \HI\ flux profiles obtained from FAST observations, in comparison to those from the V-model-flare model and the halo-inclusive model (V-model+halo), measured perpendicular to the galactic plane. We find that the \HI\ flux profiles from FAST are more similar to the results of the V-model+halo model than to those of the V-model-flare model, particularly at large galactic heights. By subtracting the data from the V-model+halo model from the FAST data and applying a $3 \times 3$ box smoothing, we also illustrate in Figure\,\ref{fig:compare-FAST_Vollmer_NGC2683} the column density maps of the excess \HI\ detected by FAST in comparison to the V-model+halo model. The column density of the excess \HI\ is predominantly distributed around $\sim 0.5 \times 10^{19}\,{\rm cm^{-2}}$, indicating that the V-model+halo model is generally consistent with the FAST observations. In Figure\,\ref{fig:residual-Channel-maps-compare}, we further present the residual channel maps of FAST data in comparison to VLA data, as well as the models V-model+halo, V-model-flare, and V-enhanced-flare. The residual maps clearly illustrate that the residuals for V-model+halo are smaller compared to those for V-model-flare and V-enhanced-flare, indicating a better alignment with FAST observations.

\begin{figure*}
\centering
	\includegraphics[width=0.85\textwidth]{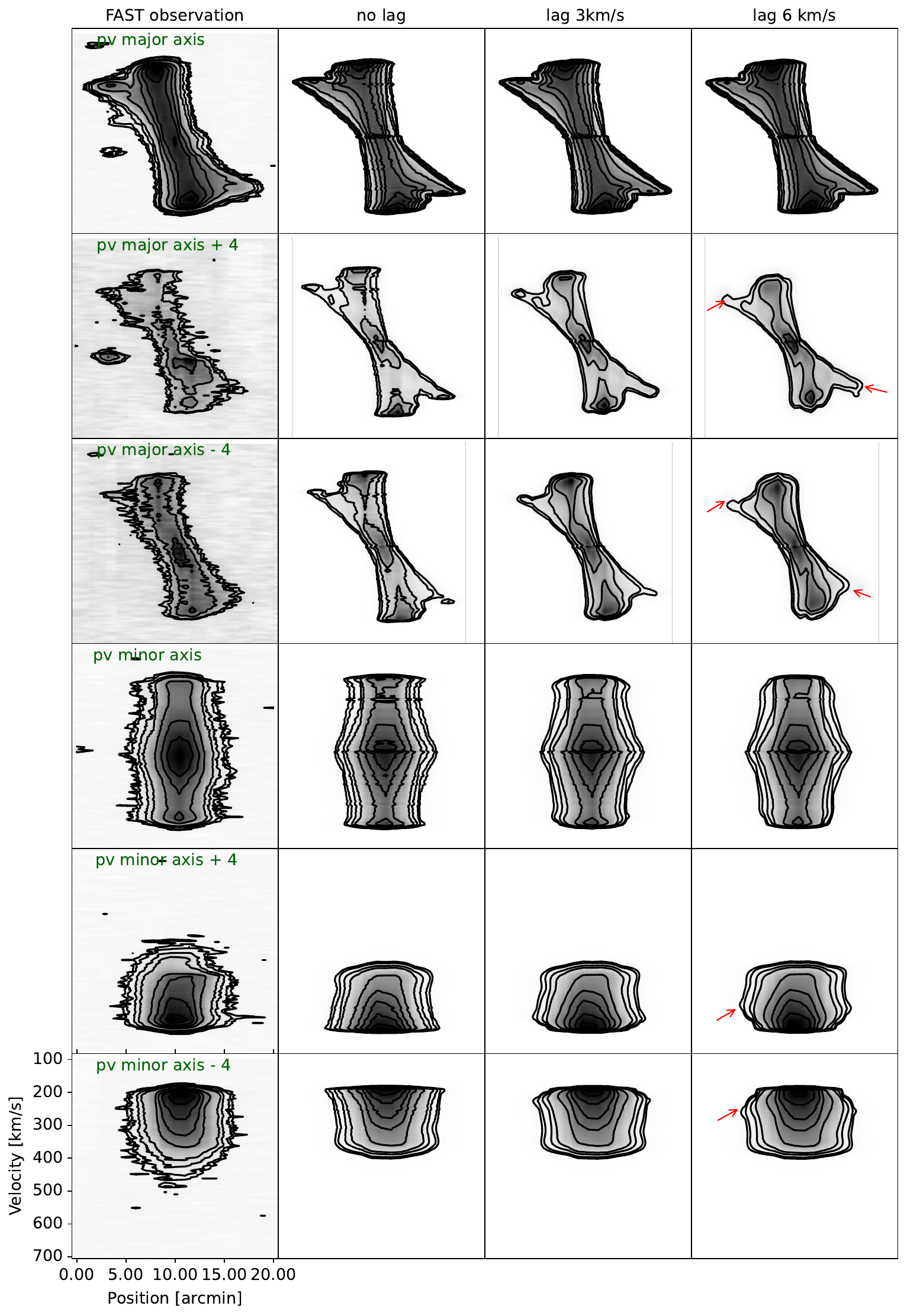}

	\caption{The \HI\ PV diagram from FAST observations is presented in the first column, alongside the \HI\ PV diagrams obtained using the V-model+halo while considering different velocity lags (the second to fourth columns correspond to lags of 0, 3 km/s, and 6 km/s, respectively). All results are shown at the resolution of FAST. The first three rows illustrate the PV diagrams along the major axis, corresponding to positions offset from the major axis by $+4'$ (toward the northeast) and $-4'$. The fourth to sixth rows depict the PV diagrams along the minor axis, corresponding to positions offset from the minor axis by $+4'$ (toward the northwest) and $-4'$. The contour levels are consistent with those in Figure\,\ref{fig:PVMaps}, and the red arrows indicate the regions of change and differences between different models and observations.}
	\label{fig:PV-maps}
\end{figure*}

In Figure\,\ref{fig:PV-maps}, we present a comparison of the PV diagrams derived from FAST observations with those from the model of V-model+halo. It is observed that in the PV diagrams along and parallel to the minor axis, the model exhibits a slightly broader range at both lower and higher velocities compared to the observations. In \citet{vollmer2015flaring}, models featuring a linear vertical gradient velocity lag of $5\,km/s$, $10\,km/s$, $15\,km/s$ were examined, and it was found that a rotation velocity lag is not a necessary component for their model. In our study, we utilized a smaller interval of $3\,km/s$ to investigate different rotation velocity lags, as shown in Figure\,\ref{fig:PV-maps}. It was found that with a velocity lag of $6\,km/s$, the PV diagrams parallel to the minor axis show a shifting of the emission distribution toward velocities closer to the systemic velocity, a phenomenon not observed in the observation. In contrast, a velocity lag of $3\,km/s$, or the absence of a velocity lag, aligns more closely with the observations. Furthermore, in the PV diagrams along and parallel to the major axis, a velocity lag of $6\,km/s$ results in a displacement of velocities in the northeast and southeast directions towards the systemic velocity, which also does not match the observational results. These findings suggest that no additional velocity lag is necessary. Additionally, we have also tested the residuals of FAST data against different velocity lag models, which similarly supports this conclusion.


\citet{marasco2019halogas} calculated the fraction of the \HI\ halo (referred to as extraplanar gas (EPG) in their article) relative to the total \HI\ mass in the HALOGAS sample. To separate the emission from the disk and EPG, they first fitted the main \HI\ disk using a Gaussian function and then constructed a cube composed of all these Gaussian fits. Subsequently, they converted the Gaussian cube into a mask by setting all voxels with intensity below (above) twice the data rms-noise to unity (blank). Applying this internal mask to the data effectively blanks all regions dominated by \HI\ emission from the regularly rotating disk, leaving the remaining \HI\ flux predominantly contributed by the EPG. Through this method, they found that the EPG component encloses approximately $5-25\%$ of the total \HI\ mass, with a mean value of 14\%. 
For NGC~2683, we employed a similar approach. We masked values in the model cube generated from the disk and flare that were below two times the rms noise of FAST, under the assumption that these regions were primarily contributed by the \HI\ halo, while values exceeding two times the rms noise were set to blank. Subsequently, we applied this masked cube to the FAST data and found that the mass of the \HI\ halo is approximately $3 \times 10^8\,{\rm M_\odot}$, which constitutes about 13\% of the total \HI\ mass. This fraction is consistent with values observed within the HALOGAS sample and is comparable to those of specific galaxies in the sample, such as NGC~5585 and NGC~4559 \citep{marasco2019halogas}, but is lower than the proportion of the \HI\ halo measured in NGC~891 \citep{oosterloo2007cold}.


The origin of the HI halo in NGC~2683 is a subject of inquiry. In the study by \citet{vollmer2015flaring}, the possibility of the galactic fountain mechanism was ruled out due to the relatively low star formation rate in NGC~2683 \citep{fraternali2006dynamical}. \citet{wiegert2015chang} utilized data from the Wide-field Infrared Survey Explorer at $22\,{\mu}$m to estimate a galaxy-averaged star formation rate (SFR) of approximately $0.09\,{\rm M_\odot \, yr^{-1}}$ for NGC~2683. Furthermore, observations of radio continuum emission with the VLA by \citet{wiegert2015chang} revealed a concentration primarily around the optical disk. The \HI\ halo we identified is notably thick, extending well beyond the optical radius distribution, presenting challenges for its formation through the galactic fountain mechanism.

The extensive vertical distribution of \HI\ is more likely attributed to external gas accretion processes. 
Our observations conducted with FAST clearly demonstrate that the dwarf galaxy dw3$?$ is positioned within the \HI\ halo in projection, with dw3$?$ interacting with the tail of KK~69 at a comparable velocity. This indicates an accretion interaction between NGC~2683 and KK~69, with dw3$?$ likely a result of such accretion. Additionally, the \HI\ emission center of dw1 is offset from its optical center, with the \HI\ center shifted towards NGC~2683. These findings collectively indicate that NGC~2683 is undergoing accretion interactions with its surrounding environment.
Furthermore, we have detected three anomalous velocity clouds, namely complexes A, B, and C. Among these, complexes B and C may act as the precursor material for the formation of the \HI\ halo, with complex B notably positioned close to dw3$?$ in projection, having already merged with the gaseous halo, indicating a high likelihood of being accreted.


\section{Conclusions}

We have conducted highly sensitive observations of the edge-on galaxy NGC~2683 and its neighboring dwarf galaxies using the FAST. Our key findings are summarized as follows:

1. Our FAST observations have unveiled a significantly more extensive \HI\ disk in the NGC~2683 galaxy compared to previous VLA observations. Particularly, perpendicular to the disk plane, we have identified an \HI\ distribution approximately four times broader than that observed by the VLA. The total \HI\ flux of the galaxy is measured at $F_{\rm HI} = 112.1\,\rm{Jy\,km\,s^{-1}}$, slightly surpassing the value of $F_{\rm HI} = 101.4\,\rm{Jy\,km\,s^{-1}}$ reported by \citet{vollmer2015flaring} based on VLA observations. Assuming a distance of 9.36$\,$Mpc, this corresponds to a total \HI\ mass of $M_{\rm HI} = 2.32 \times 10^9\,{\rm M_\odot}$.

2. FAST has detected three dwarf galaxies near NGC~2683: KK~69, dw1, and dw3$?$ with significantly enhanced sensitivity compared to previous observations. Channel maps indicate a subtle link between the tail of KK~69 and dw3$?$, which lacks an optical counterpart. This suggests that dw3? may result from accretion originating from the dwarf galaxy KK~69 towards NGC~2683.  We computed the flux and mass values for these three dwarf galaxies. Furthermore, FAST has identified three compact HVCs, namely complexes A, B, and C. Complex A is more likely to be a Galactic HVC, whereas complexes B and C are more likely associated with the NGC~2683 group. Additionally, we found that complex B appears to be in the process of being accreted onto NGC 2683. We also determined the mass of the three complex clouds by assuming them as the HVCs of the NGC~2683 group or the UCHVCs of the local group, respectively.

3.  By utilizing the foundational model from \citet{vollmer2015flaring} and incorporating the \HI\ halo component, we found that the model with the added \HI\ halo provides a better explanation for our observations from FAST. The estimated mass of the \HI\ halo in NGC~2683 is $3 \times 10^8\,{\rm M_\odot}$, accounting for approximately 13\% of the total \HI\ mass of the galaxy. The presence of such an extensive \HI\ halo is more likely due to accretion from neighboring dwarf galaxies and HVCs. 

\begin{acknowledgments}

We are deeply grateful to J. Saponara for contributing the GMRT data of KK~69 and the VLA data of dw1 and dw3$?$.
This work is supported by National Natural Science Foundation of China (NSFC, Nos. 12033004, 12003070, and 12233005), Research Funding of Wuhan Polytechnic University NO. 2022RZ035. Y. Gao acknowledges support from Scientific Research Fund of Dezhou University, 3012304024, and Shandong Provincial Natural Science Foundation (ZR2024QA212).

\end{acknowledgments}

\bibliography{NGC2683_references}{}
\bibliographystyle{aasjournal}

\end{document}